\documentclass[twocolumn,prb,superscriptaddress,longbibliography]{revtex4-2}

\usepackage{amsmath,amsfonts,amssymb}
\usepackage{graphicx,color}
\usepackage{hyperref}
\usepackage{epstopdf}
\usepackage{float}
\usepackage{multirow}
\usepackage{hhline}
\usepackage{ulem}
\usepackage{mathtools}
\usepackage[caption=false]{subfig}
\usepackage{soul}

\usepackage{graphicx}
\usepackage{dcolumn}
\usepackage{bm}

\usepackage[utf8]{inputenc} 

\usepackage{multirow}
\usepackage[warn]{mathtext}
\usepackage{amssymb}

\usepackage{textcomp} 
\usepackage{indentfirst} 
\usepackage{amsmath} 
\usepackage{graphicx}
\DeclareGraphicsExtensions{.pdf,.png,.jpg}
\usepackage{pgfplots}
\pgfplotsset{compat=1.13}

\usepackage{hyperref}
\hypersetup{
    colorlinks=true,
    linkcolor=blue,
    filecolor=black,      
    urlcolor=blue,
    citecolor= violet
}

\usepackage{algpseudocode}

\usepackage{adjustbox}
\usepackage{tabularx}

\usepackage{mathtools}

\usepackage{xcolor}
\definecolor{C0}{HTML}{4C72B0}
\definecolor{C1}{HTML}{DD8452}
\definecolor{C2}{HTML}{55A868}
\definecolor{C3}{HTML}{C44E52}

\newcommand{\ilyas}[1]{\textcolor{black}{#1}}
\newcommand{\lena}[1]{\textcolor{black}{#1}}



\newcolumntype{Y}{>{\centering\arraybackslash}X}

\makeatletter
\def\@fnsymbol#1{\ensuremath{\ifcase#1\or \dagger\or \ddagger\or
   \mathsection\or \mathparagraph\or \|\or **\or \dagger\dagger
   \or \ddagger\ddagger \else\@ctrerr\fi}}
\makeatother

\begin{document}


\title{\lena{Three-mode tunable coupler for superconducting two-qubit gates }}

\author{Elena~Yu.~Egorova}
\email[Corresponding author: ]{yelena.egorova@phystech.edu}
\affiliation{National University of Science and Technology MISIS, 119049 Moscow, Russia}
\affiliation{Russian Quantum Center, 143025 Skolkovo, Moscow, Russia}
 \affiliation{Moscow Institute of Physics and Technology, Dolgoprudny 141700, Russia}

\author{Alena~S.~Kazmina}
\affiliation{National University of Science and Technology MISIS, 119049 Moscow, Russia}
\affiliation{Russian Quantum Center, 143025 Skolkovo, Moscow, Russia}
 \affiliation{Moscow Institute of Physics and Technology, Dolgoprudny 141700, Russia}

\author{Ilya~A.~Simakov}
\affiliation{National University of Science and Technology MISIS, 119049 Moscow, Russia}
\affiliation{Russian Quantum Center, 143025 Skolkovo, Moscow, Russia}
 \affiliation{Moscow Institute of Physics and Technology, Dolgoprudny 141700, Russia}

\author{Ilya~N.~Moskalenko}
\affiliation{National University of Science and Technology MISIS, 119049 Moscow, Russia}

\author{Nikolay~N.~Abramov}
\affiliation{National University of Science and Technology MISIS, 119049 Moscow, Russia}

\author{Daria~A.~Kalacheva}
\affiliation{Skolkovo Institute of Science and Technology, Skolkovo Innovation Center, Moscow 121205, Russia}
\affiliation{Moscow Institute of Physics and Technology, Dolgoprudny 141700, Russia}
\affiliation{National University of Science and Technology MISIS, 119049 Moscow, Russia}

\author{Viktor~B.~Lubsanov}
\affiliation{Moscow Institute of Physics and Technology, Dolgoprudny 141700, Russia}

\author{Alexey~N.~Bolgar}
 \affiliation{Moscow Institute of Physics and Technology, Dolgoprudny 141700, Russia}
 \affiliation{Skolkovo Institute of Science and Technology, Skolkovo Innovation Center, Moscow 121205, Russia}
\affiliation{Russian Quantum Center, 143025 Skolkovo, Moscow, Russia}

\author{Nataliya~Maleeva}
\affiliation{National University of Science and Technology MISIS, 119049 Moscow, Russia}

\author{Ilya~S.~Besedin}
\thanks{Present Address: Department of Physics, ETH Zurich, Zurich, Switzerland}
\affiliation{National University of Science and Technology MISIS, 119049 Moscow, Russia}
\affiliation{Russian Quantum Center, 143025 Skolkovo, Moscow, Russia}

\date{\today}

\begin{abstract}



Building a scalable universal high-performance quantum processor is a formidable challenge.
In particular, the problem of realizing fast high-perfomance two-qubit gates of high-fidelity remains needful. Here we propose a \lena{building block} for a scalable quantum processor consisting of two transmons and a tunable three-mode coupler allowing for a ZZ interaction control.
We experimentally demonstrate the native CZ gate with the pulse duration of 60~ns achieving the two-qubit gate fidelity above 98$\%$, limited mostly by qubit coherence time. Numerical simulations show that by optimizing the gate duration the fidelity can be pushed over $99.97\%$.

\end{abstract}

\maketitle


\section{\label{sec:introduction}Introduction}

Development of superconducting qubit-based quantum computing devices has achieved a number of major milestones during the past few years \cite{Arute2019, Wu2021, chen2021exponentia, google2023suppressing, kim2023evidence}. The common theme is the usage of transmon qubits \cite{Koch2007} with dispersive readout and capacitive coupling \cite{Blais2004}. Nevertheless, the relatively large two-qubit gate error rates \lena{remain to be the limiting factor for} both the practical application of such machines as well as the proof-of-principle advantage demonstrations \cite{Pan2022}.

\lena{For transmon qubits, there is variety of different implementations of two-qubit gates}, each with its strengths and weaknesses. The simplest scheme involves resonant exchange of qubit populations triggered by a frequency shift of one of the qubits \cite{Pashkin2003, Strauch2003}. For transmons, one can choose between an iSWAP-type gate \cite{Dewes2012, PhysRevLett.125.120504, Ganzhorn2020} or a CPHASE-type gate \cite{PhysRevLett.125.120504, Ganzhorn2020, DiCarlo2009}. The gate duration can be short and is determined by the qubit-qubit coupling constant. This scheme is prone to (i) significant qubit dephasing, because the qubits are operated outside of their sweet spots, which can be addressed with net-zero pulses \cite{Negirneac2021}, (ii) crossing of two-level system (TLS) defects during gate operation \cite{Krinner2022}, and (iii) large residual ZZ crosstalk when the qubits are detuned from each other. The need to traverse a large frequency range with the qubit and thus potentially cross strongly coupled TLS defects can be circumvented, for example, using a parametrically modulated signal \cite{Caldwell2018}. For parametrically activated gates, however, the effective coupling strength is only a fraction of the underlying coupling between qubits. Moreover, the qubit is still detuned from its sweet spot, resulting in additional dephasing, which can be reduced by applying a modulation around the DC sweet spot \cite{Hong2020}. 

Another approach is \lena{based on} tunable coupler elements, either for the parametric modulation gate \cite{Niskanen2007, mckay2016universal} or direct resonance gate \cite{Li2020, PhysRevX.13.031035}. For CPHASE-type gates one can also use an adiabatic flux pulse, which results in slower gates for the same coupling strength \cite{Collodo2020, PhysRevLett.127.080505, Kubo_2023, kubo2024highperformance, li2024realization}, but does not require precise resonance between the qubits. Otherwise, one must tune both the qubit frequency and the coupler. To avoid dephasing, one can use an AC signal to tune the effective qubit frequency for the duration of the gate \cite{Sete2021}. By controlling both qubit frequencies and coupler flux with DC signals, one can implement the full continuous gate set that, up to single-qubit gates, includes all excitation-number-preserving unitaries \cite{PhysRevLett.125.120504}. For tunable couplers, one of the main desired features is cancelling out crosstalk. This can be done in a straightforward manner by using destructive interference between two coupling paths. These coupling paths can either be of Fano-type, with one coupling being resonant \cite{Yan2018, jin2024fasttunablehighfidelity, roy2022realization} or both coupling paths having resonances \cite{Mundada2019}. In either case, the residual interaction strongly depends on the relation between qubit and coupler frequencies.

To address the issue of a relatively large residual ZZ interaction, several techniques can be identified capable of mitigating this unwanted effect. Among the most common methods are passive suppression using highly detuned qubits \cite{Collodo2020, Xu2020} or possessing opposite-sign anharmonicities \cite{Zhao_2020, Ku_20}, as well as active dynamical decoupling protocols \cite{Viola98, Tripathi22}, which are widely employed in experiments involving long idling times \cite{chen2021exponentia, google2023suppressing}. Another promising strategy involves utilizing the AC Stark shift effect, which has demonstrated encouraging performance in recent experiments~\cite{Mitchell_2021, Noguchi_2020}. Alternatively, similar suppression can be achieved by applying a weak microwave drive to the coupler~\cite{Ni_2022}.
\color{black}

In this work, we present a \lena{building block} suitable for a scalable quantum processor architecture and consisting of a two concentric transmons \cite{Braumuller2016, Rahamim2017, Caldwell2018, Martinis2022, Eun2022} and a tunable three-mode coupler for a controlled ZZ interaction.
The distinctive feature of the three-mode coupler is its lack of dependence on high asymmetry in the Josephson junctions of the coupler SQUID. This characteristic provides a significant advantage in cases where achieving high reproducibility of junctions with different sizes is challenging. For instance, in the case of the \lena{double-transmon} coupler type \cite{li2024realization}, the achieved asymmetry is critical, as deviations can lead to significant changes in the energy structure of the flux qubit. Similarly, a transmon-type coupler \cite{Collodo2020, PhysRevLett.127.080505, Yan2018, Mundada2019} requires precise control of asymmetry to simultaneously achieve both a small residual ZZ interaction and a coupling strength on the order of tens of megahertz at the operating point. The proposed three-mode coupler achieves comparable levels of residual ZZ interaction strength while providing a significantly larger tunable range compared to conventional single-mode couplers, assuming identical design parameters.
By tuning the magnetic flux in the coupler SQUID, we are able to vary the interaction strength between qubits by two orders of magnitude. 
We implement a native two-qubit CZ gate by leveraging a cross-Kerr interaction \cite{Collodo2020, PhysRevLett.127.080505, li2024realization, PRXQuantum.3.020301, PRXQuantum.4.010314}
and estimate a gate fidelity above 98\% using cross-entropy benchmarking, limited mostly by the qubits' coherence time.

\section{\label{sec:Device_description}Device description}
\begin{figure}[H]
    \centering
    \normalsize
    \includegraphics[width=1\columnwidth]{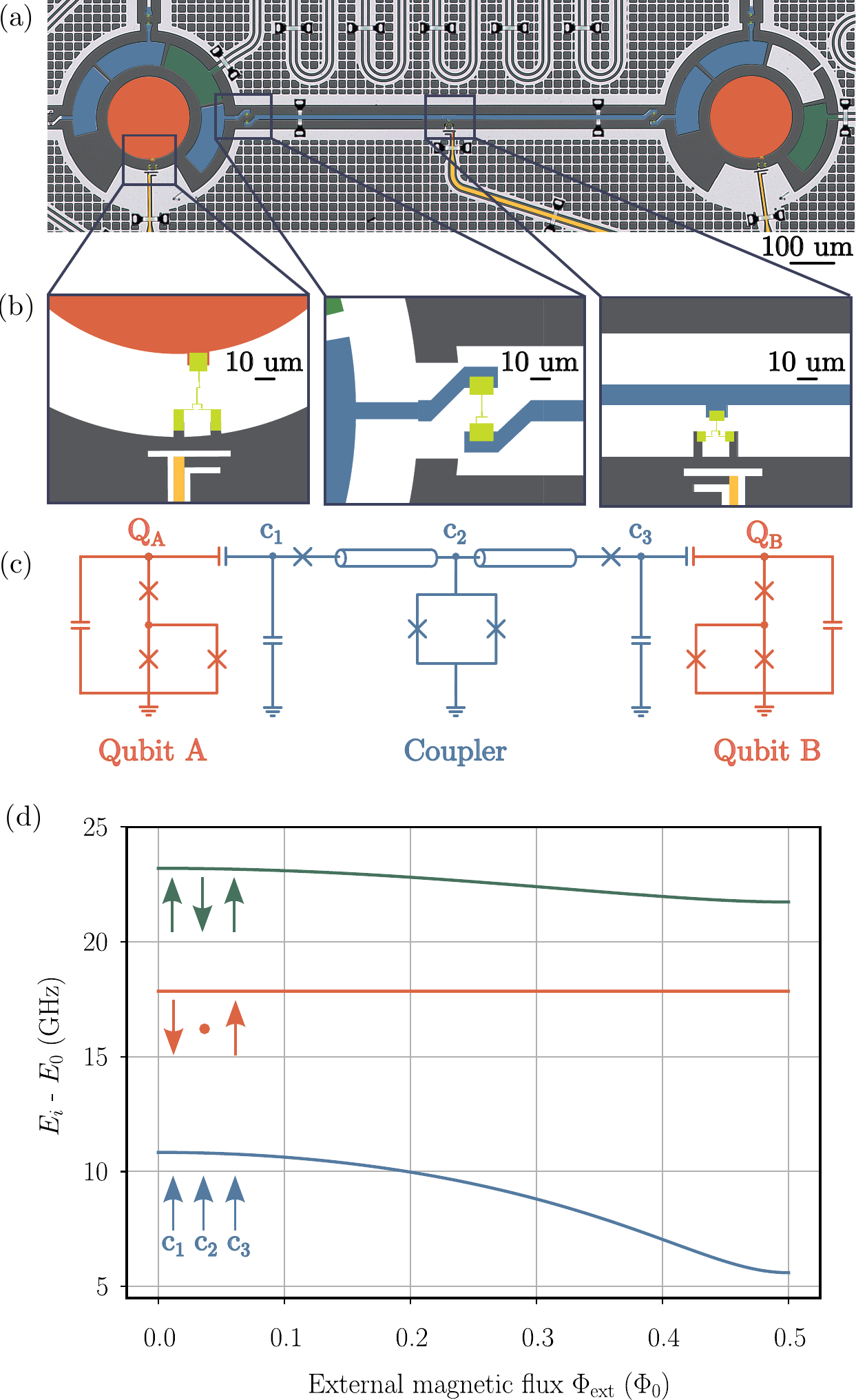}
	\caption{The two-qubit cell.
 (a) Optical image of \lena{the} sample. Each transmon (red) has a round island  grounded through a circuit of three Josephson junctions: \lena{a single} junction in series with a SQUID (yellow). Each qubit has a separate \lena{flux} control line both for \lena{frequency} tuning and excitation. The transmons  are connected to each other through a tunable coupler (blue). The coupler is realised by a coplanar line with a SQUID in the middle. \lena{At either end of the coupler, there is an arc-shaped electrode with a large mutual capacitance with the respective qubit island.} A separate line controls the magnetic flux through the coupler's SQUID, providing frequency tunability. Additional Josephson junctions are embedded at each end of the coplanar. 
(b) Three key elements of the cell (from left to right): the three Josephson junction circuit of the transmon, the Josephson junction at one of the coplanar ends, the SQUID of the coupler. 
(c) \lena{Simplified} electrical circuit \lena{schematic} of the two-qubit cell. The two transmons are shown in red, the coupler 
is shown in blue. The more detailed electrical circuit can be found in Appendix~\ref{sec:Appendix1}.
(d) Numerically calculated coupler energy levels as a function of the external magnetic flux in the coupler SQUID (see Appendix~\ref{sec:Appendix1}). The normal modes of the coupler are determined by the states of its three nodes. Arrows indicate \lena{signs of} charge oscillations across these nodes for each mode. 
}
    \label{fig:qcq_zoom}
\end{figure}

\lena{The two-qubit device consists} of two tunable transmons connected by a tunable three-mode coupler with an individual flux-control line. We utilize concentric weakly flux-tunable transmon qubits \cite{egorova2024} for reduction of dielectric losses, containing a \lena{three-Josephson-junction (3-JJ)} circuit \cite{ChavezGarcia2022} for reducing flux noise sensitivity at the expence of narrowing tunability range, shown in Fig.~\ref{fig:qcq_zoom}. Our 3-JJ qubit circuit consists of two \lena{Josephson junctions (JJs)} forming an asymmetric DC SQUID connected in series to a third JJ. Each qubit has a single flux-coupled control line for both dc and microwave \lena{signals}. The physical pitch between qubits is 2~mm. Capacitive coupling between the transmon and the coupler is provided by arc-shaped electrodes \cite{Caldwell2018, andersen2020repeated}. The three-mode coupler has three nodes and consists of a coplanar waveguide segment, terminated with JJs, and shorted in the middle by a DC SQUID, see~Fig.~\ref{fig:qcq_zoom}(b). \lena{The Hamiltonian of the system and the circuit parameters are described in Appendix A. Due to the mirror-symmetry of the circuit, the eigenmodes are either even or odd (Fig.~\ref{fig:qcq_zoom}(d)). The even modes are flux-tunable. The odd mode is not flux-tunable, since there is no currents flowing through the SQUID associated with it.}

\lena{Al/AlOx/Al Josephson junctions in the device are formed using a standard Dolan bridge technique \cite{dolan1977offset} in a single vacuum cycle, ensuring they share the same critical current density. The galvanic contact with the ground plane is achieved by depositing bandages through a single-layer organic mask following argon milling of the aluminum oxide \cite{osman2021simplified}. More details on the fabrication can be found in \cite{kazmina2024demonstration}.}


\begin{figure*}
    \centering
    \includegraphics[width=1.\linewidth]{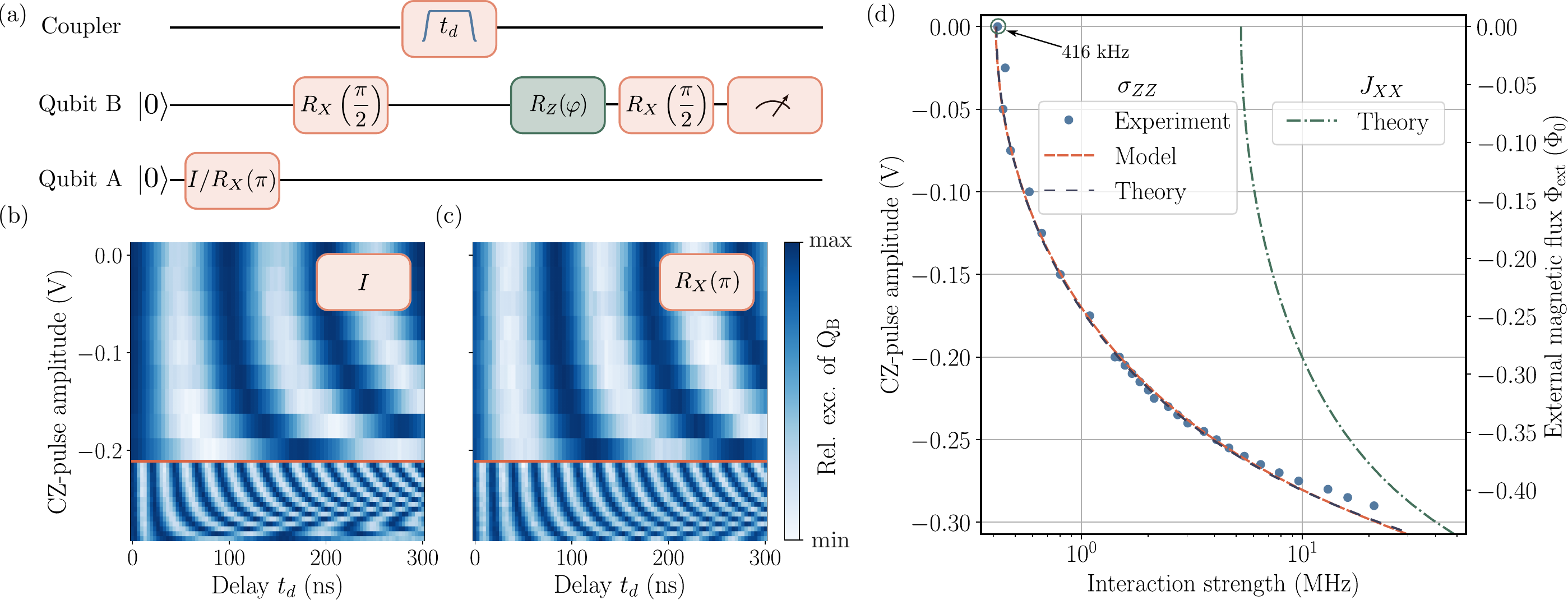}
    \caption{
    The measurement of the ZZ interaction. 
    (a) The pulse sequence performed on our device: a Ramsey-type experiment on the qubit B while control qubit A is initialized in its ground or excited state, the coupler is tuned by a Hann-shaped flux pulse of the duration $t_d$. 
The argument $\varphi$ of the virtual $R_Z$ gate is determined by an offset frequency $f_\text{offset}$ and pulse duration $t_d$.
    (b, c) The resulted pattern of the qubit B relative excitation. 
    The red line indicates the boundary between two different offset frequencies used in the experiments. 
    (d) The ZZ interaction strength defined as the difference between the oscillation frequencies of the qubit B population in (b) and (c) (blue dots).
    The red dashed line shows the ZZ interaction strength $\zeta_{ZZ}$, calculated using the full-circuit model described in Appendix~\ref{sec:Appendix1} with the experimental parameters obtained by two-tone spectroscopy measurements of each computational qubit and the coupler. 
    The optimized parameters are given in Appendix~\ref{sec:Appendix1} Table~\ref{tab:params}. The blue dashed and green dashed-dotted lines represent the ZZ interaction strengths, \(\zeta_{ZZ}\) and the effective XX coupling, $J_\mathrm{XX}$, calculated using the approach described in Section~\ref{sec:zz_int} and Appendix~\ref{sec:Appendix_ZZ}, with the experimental parameters as well. \color{black}The idle bias point corresponds to the zero external magnetic flux, where the smallest ZZ interaction between qubits is achieved.
    }
    \label{fig:zz_calibration}
\end{figure*}

\section{\label{sec:results}Results}
\subsection{\label{sec:ZZ_interaction}ZZ interaction strength}

In our experiment we measure the strength of the ZZ interaction between qubits as a function of an external flux through the coupler SQUID.
We use a Ramsey-type pulse sequence, consisting of two $\pi/2$ pulses applied to the qubit B, and a Hann-shaped flux pulse of duration $t_d$, applied to the coupler, between them, as depicted in Fig.~\ref{fig:zz_calibration}(a). \lena{The results of the Ramsey experiment on qubit B as a function of flux pulse amplitude and length are shown in Fig.~\ref{fig:zz_calibration}(b) for de-excited qubit A and in Fig.~\ref{fig:zz_calibration}(c) for excited qubit A.}
For each voltage amplitude generating a flux through coupler SQUID, we find the frequency of population oscillations as a function of the delay $t_d$.
We define the strength of the ZZ interaction $\zeta_{ZZ}$ as the difference of these frequencies.
In order to observe oscillations on scales of hundreds of nanoseconds even at low external flux, we apply a virtual rotation $R_Z(\varphi)$ \cite{mckay2017efficient} about the Z axis on qubit B before the second $R_X(\pi/2)$ gate.
The phase $\varphi$ depends on the duration of the coupler pulse $\varphi = 2\pi f_\text{offset} t_d$, where $f_\text{offset}$ is a defined offset from the frequency of the qubit B.
For convenience, we take two offset frequencies equal to 10 MHz for low and 50 MHz for high CZ-pulse amplitudes.
The red line in Fig.~\ref{fig:zz_calibration}(b, c) \lena{shows the boundary between the measurements with the different offset frequencies}.

The measured ZZ interaction strength \lena{are shown} in Fig.~\ref{fig:zz_calibration}(d) \lena{with} blue points. 
We extract system parameters from the two-tone spectroscopy of both qubits. Knowing the qubit frequencies $f_\mathrm{q_A}, f_\mathrm{q_B}$ and anharmonicities $\delta_\mathrm{q_A}, \delta_\mathrm{q_B}$ \color{black}as functions of qubit external flux, \lena{we estimate the} critical currents \lena{of the six Josephson junctions} of the qubit \lena{circuits with least-squares fitting}.
From the two-tone spectroscopy of the coupler we obtain the coupler frequency $f_\text{C}$ at external magnetic flux $\Phi_\mathrm{ext} = 0.5\Phi_0$. 
Using the full system Hamiltonian with obtained qubit JJ \lena{critical currents} and capacitances from the design, we \lena{estimate the critical currents of} the four JJs related to the coupler, colored blue in Fig.~\ref{fig:qcq_zoom}(c).
\lena{In the detailed scheme in Fig.~\ref{fig:electrical scheme} these junctions are noted as $J_5, J_6$ and $J_4=J_7$. The last two junctions are considered to be equal.}
We optimize the three critical currents of the coupler junctions in order to meet an experimental values of the ZZ interaction at the zero external magnetic flux $\zeta_{ZZ}(\Phi_\text{ext}=0)$ and the coupler frequency $f_\text{C}$.
The designed and fitted parameters \lena{of} the system Hamiltonian (\ref{full_ham}) are presented in \lena{Tab.~\ref{tab:params}}. 
In the optimization procedure, we vary only the JJs areas, assuming the same critical current density for all of them. 

During the study of the ZZ interaction between qubits, both transmons are kept in the upper flux sweet spot, while the magnetic flux in the coupler loop is varied between zero flux and half flux quantum $\Phi_\mathrm{ext} \in \left[0, \Phi_0/2\right]$.
In each subsystem, we take into account the first five energy levels. 
The ZZ interaction strength, $\zeta_{ZZ}$, is calculated by the system Hamiltonian~\eqref{full_ham} numerical diagonalization and  is defined as the frequency difference between the state in which both computational qubits are excited $|ee\rangle$ and the sum of the frequencies corresponding to the states where only one qubit is excited, $|eg\rangle$ and $|ge\rangle$: $\zeta_{ZZ} = f_{ee} - f_{eg} - f_{ge}$, assuming the ground state frequency is set to zero.
\color{black}
The simulated dependence of $\zeta_{ZZ}$ on the external magnetic flux $\Phi_\text{ext}$ is shown in Fig.~\ref{fig:zz_calibration}(d), together with the experimental data.
We denote the states of the computational qubits as $|g\rangle, |e\rangle, |f\rangle$ and use numerical notation for coupler modes. For simplicity, we omit explicit indices for the coupler state and assume projection onto its ground state, for instance, the state \(|ee\rangle\) corresponds to \(|ee000\rangle\). 
Details on the numerical calculation can be found in Appendix~\ref{sec:Appendix1}.

At the idle bias point of the coupler ($\Phi_\mathrm{ext} = 0$), we obtain a residual ZZ interaction constant of 416 kHz. This by far exceeds the target value of 146 kHz obtained from calculations with the design parameters (see Fig.~\ref{fig:spectrum}(b) in Appendix~\ref{sec:Appendix1}). 
Such a difference in $\zeta_\mathrm{ZZ}$ is caused by fabrication imperfections, i.e. by the difference between the design and the fabricated device. However, the difference between expected and approximated parameters does not significantly influence the overall gate performance, as we show in further simulations and experiments.

\subsection{\label{sec:zz_int}Linear-response theory for qubit-qubit coupling}
In the linear response regime, it is possible to derive an analytical expression for the XX interaction strength between the qubits. Given the capacitance matrix and the diagonal inverse inductive matrix of the full system, the dependence of the coupler node fluxes on the qubit current $\tilde{I}_\mathrm{q_A}$ can be determined. The voltage of the second qubit $\tilde{V}_\mathrm{q_B}$ is then expressed in terms of the coupler fluxes, allowing the construction of the impedance between the qubits,  $Z_\mathrm{q_A q_B} = \tilde{V}_\mathrm{q_B}/\tilde{I}_\mathrm{q_A}$. Following the methodology outlined in \cite{impedance, solgun2022direct}, the XX coupling is defined as a function of the impedance between qubits:
\begin{equation}
\begin{split}
J_\mathrm{q_A q_B} = -\frac{1}{4}\sqrt{\frac{\omega_\mathrm{q_A}\omega_\mathrm{q_B}}{L_\mathrm{q_A}L_\mathrm{q_B}}}\textnormal{Im}\left[\frac{Z_\mathrm{q_A q_B}(\omega_\mathrm{q_A})}{\omega_\mathrm{q_A}} + \frac{Z_\mathrm{q_A q_B}(\omega_\mathrm{q_B})}{\omega_\mathrm{q_B}}\right],
\end{split}
\end{equation}
where $\omega_\mathrm{q_A}$ and $\omega_\mathrm{q_B}$ are qubit mode frequencies, $L_\mathrm{q_A}$ and $L_\mathrm{q_B}$ are the total inductances of 3-JJ qubit circuits in this case.\color{black} The total XX coupling between qubits is estimated as $J_{XX} = J_\mathrm{q_A q_B} + J_0$, where $J_0$ represents the direct capacitive interaction between the qubits. 
The imaginary part of the impedance here is
\begin{equation}
\begin{split}
 \mathrm{Im}Z_\mathrm{q_A q_B}(\omega)= \omega\left(\frac{C_\textnormal{23}}{C_\textnormal{23}+C_\mathrm{q_A}}\right)^2 \times \\\times\frac{L_\textnormal{J56}\omega^2_\mathrm{c_+}\omega^2_\mathrm{c_0}\omega^2_\mathrm{c_-}}{(\omega^2_\mathrm{c_+}-\omega^2)(\omega^2_\mathrm{c_0}-\omega^2)(\omega^2_\mathrm{c_-}-\omega^2)},
\end{split}
\end{equation}
where \lena{$L_{\mathrm{J}56}$ is the flux-dependent effective inductance of the coupler SQUID, and} $\omega_{c_+}, \omega_{c_0}, \omega_{c_-}$ are coupler normal modes frequencies depicted in Fig.~\ref{fig:qcq_zoom}(d). The mode indices indicate the relative directions of charge oscillations at the central coupler node with respect to the side nodes. Further details can be found in the Appendix~\ref{sec:Appendix_ZZ}.

Once the XX coupling is known (see the green dashed-dotted line in Fig.~\ref{fig:zz_calibration}(d)), \color{black} the ZZ interaction strength can be determined. The Hamiltonian of the two-qubit system is constructed in the model of Duffing oscillators:
\begin{equation}
    \begin{split}
        H &= \omega_\mathrm{q_A} a ^ \dagger a + \frac{\delta_\mathrm{q_A}}{2} a ^ \dagger a  (a ^ \dagger a  - 1) 
        +\omega_\mathrm{q_B} b ^ \dagger b +\\ &+ \frac{\delta_\mathrm{q_B}}{2} b ^ \dagger b  (b ^ \dagger b  - 1) + J_{XX} (a ^ \dagger b + a b ^ \dagger).
    \end{split}
\end{equation}
The numerical Hamiltonian is constructed using the energy levels $|fg\rangle$, $|gf\rangle$, and $|ee\rangle$, while accounting for the XX interaction between levels. The ZZ interaction strength is determined as the frequency difference between the $|ee\rangle$ energy level in the undressed and dressed Hamiltonians. 
The results of this calculation are compared with experimental data and numerical approach in Fig. \ref{fig:zz_calibration}(d). Additional details are provided in Appendix~\ref{sec:Appendix_ZZ}. A comparison between the proposed coupler and the single-mode tunable coupler is presented in Appendix~\ref{sec:AppendixComp}. An analysis of the scheme involving highly detuned qubits mediated by the three-mode coupler is provided in Appendix~\ref{sec:AppendixDetuned}.
\color{black}

\subsection{\label{sec:Gate_concept}CZ gate concept}

\begin{figure*}
    \centering
    \includegraphics[width=1.\linewidth]{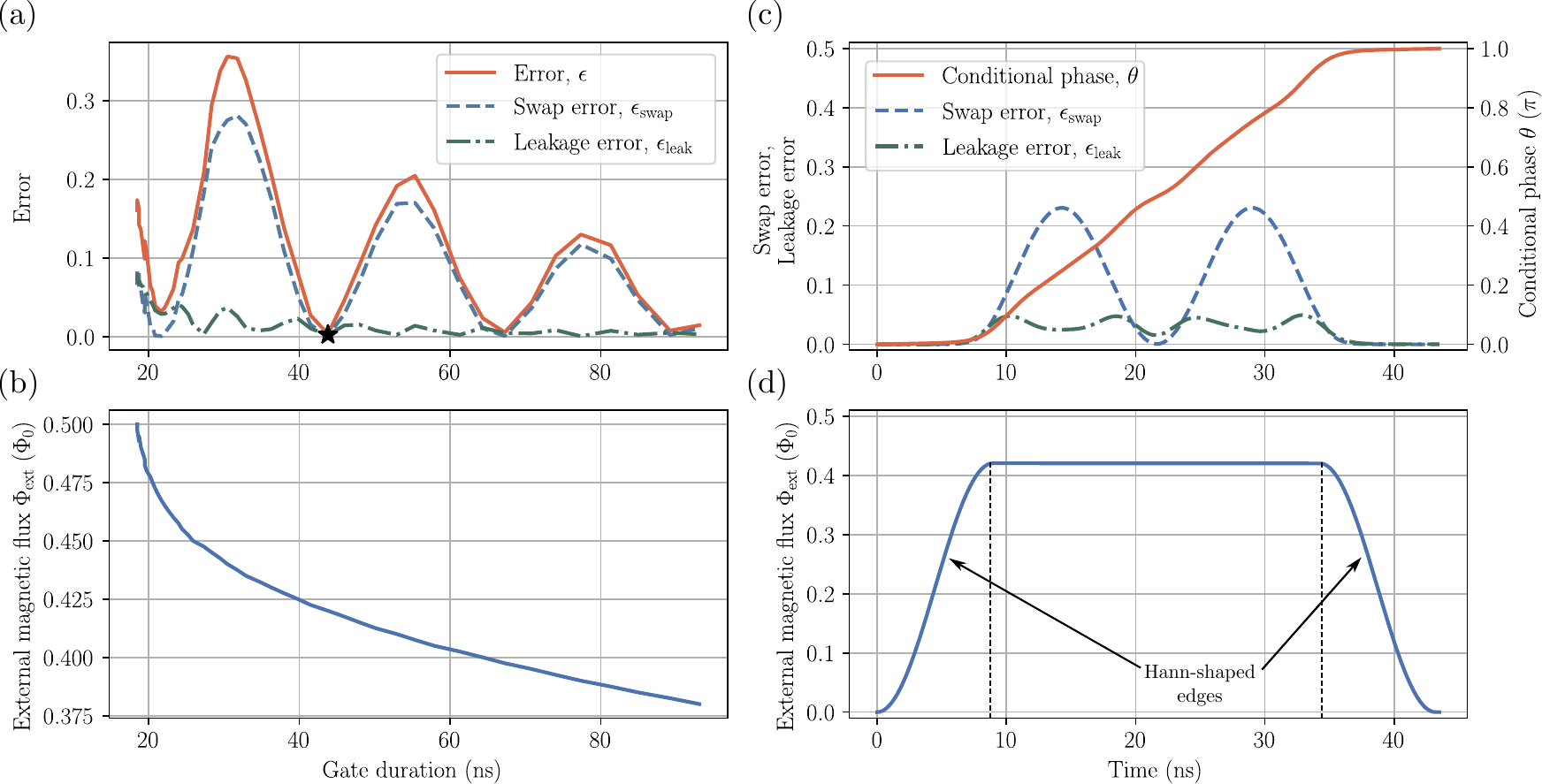}
    \caption{The CZ gate simulation for the experimentally obtained device parameters.
    (a) The error (red), the swap error $\epsilon_\mathrm{swap}$ between the computational states (dashed blue) and the high-energy levels leakage error $\epsilon_\mathrm{leak}$ (dashed-dotted green) are shown as functions of the CZ gate duration. 
    (b) The corresponding amplitude of the external flux in the coupler's SQUID. 
    (c) For the gate duration of 43.3 ns (star in panel (a)) the time evolution of the conditional two-qubit phase $\theta$ (red), the swap error $\epsilon_\mathrm{swap}$ (dashed blue), and the leakage error $\epsilon_\mathrm{leak}$ (dashed-dotted green) are simulated.
    The estimated fidelity estimated is $F =0.9997$. 
    (d) The coupler flux pulse with Hann-shaped edges are used both in simulation and experiment.
    }
    \label{fig:pulse}
\end{figure*}

\lena{We realize the two-qubit CZ gate with a flux pulse applied to the coupler SQUID.}
One can switch on the interaction adiabatically increasing the external flux and, after waiting for the required time, return the coupler to the idle bias point. The flux pulse \lena{should be as short as possible to avoid decoherence during the gate} while remaining adiabatic in order to prevent undesired transitions between the computational states and leakage out of the computational subspace.
Therefore, in order to activate the interaction we use a smooth pulse $\Phi_\text{ext}(t)$ with a Hann function-shaped edges of 9~ns both in the simulation and the experiment, as it is shown in Fig.~\ref{fig:pulse}(d).
Also, we define the gate duration as the length of the entire pulse, accounting \lena{for} both edges and a plateau-shaped part.

We described above in details the method of constructing the system model using two-tone spectroscopy data and measurement of the ZZ interaction strength at the zero-flux point.
Using the obtained values, we construct the full-circuit Hamiltonian and calculate the time dynamics of the system during the CZ gate implementation.
In this numerical analysis, we take into account the first four energy levels of each qubit and each mode of the coupler in order to reduce a computational time.
Thus, the Hilbert space dimension of the full system in the simulation is $4^5$. 


We simulate the CZ gate performance for a range of the coupler's SQUID flux pulse amplitude from $0.37\Phi_0$ to $0.5\Phi_0$.
For each amplitude value we find a gate duration, such that the conditional phase accumulated by the computational state $|ee\rangle$ is $\theta = \pi$.
In order to evaluate the performance of the resulting CZ operation, we use the standard expression for the two-qubit gate fidelity \ilyas{utilizing Pauli transfer matrix representation \cite{nielsen2002simple, PhysRevLett109060501}, which we then convert to the depolarizing error $\epsilon$.}
To analyze the error contribution, we calculate the average leakage error out of the computational subspace \cite{Wood_2018}
\begin{equation}  
    \epsilon_\mathrm{leak} = 1 - \frac{1}{4} \sum_{i,j=1}^4 \left| u_{ij} \right|^2,
\end{equation} 
and the swap error, \(\epsilon_\mathrm{swap}\), which quantifies undesired for a CZ gate state population exchange within the computational subspace 
\begin{equation}  
    \epsilon_\mathrm{swap}  = \frac{1}{4} \sum_{\substack{i,j=1\\i \neq j}}^4 \left| u_{ij} \right|^2.
\end{equation}  
Here, $u_{ij}$ denotes the matrix elements of the evolution operator $\hat{\mathcal{U}}$ projected onto the two-qubit computational subspace, with the coupler in the ground state.

\ilyas{The resulting errors are} presented in Fig.~\ref{fig:pulse}(a), while the corresponding gate duration as a function of an external flux amplitude is given in Fig.~\ref{fig:pulse}(b).
Here we observe a periodic behavior of \ilyas{the gate error condutioned by population oscallations within the computational subspace.} 
At the first minimum of the fidelity curve we do not get high-fidelity operation due to the immense ZZ interaction.

\begin{figure*}
    \includegraphics[width=1.\textwidth]{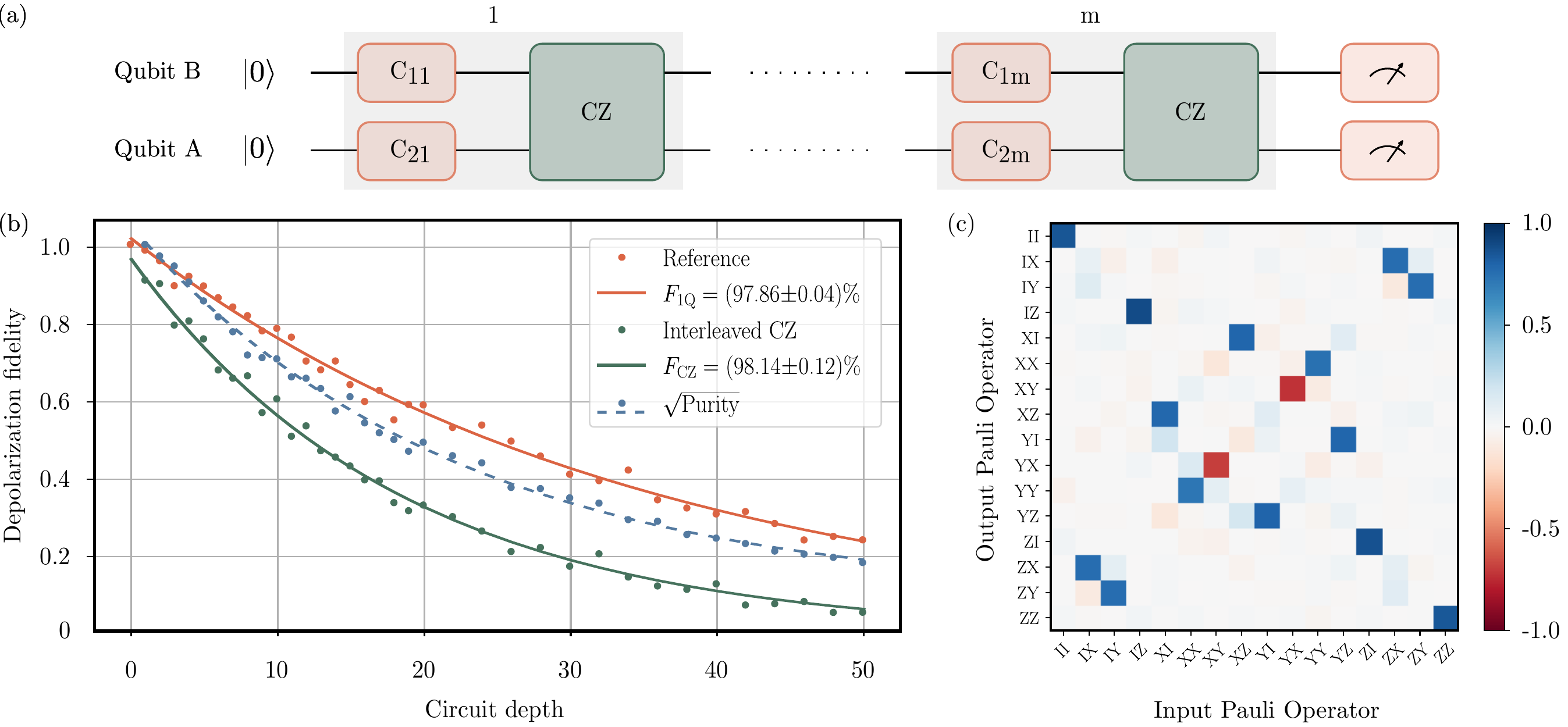} 
	\caption{The experimental benchmarking of the CZ gate. (a) The pulse sequence for the cross-entropy benchmarking (XEB) experiment with an interleaved CZ gate. (b) The depolarization fidelity averaged over 100 random circuits as a function of the circuit depth $m$ for the reference sequence: sequence (a) without CZ gates
 (red) and with an interleaved CZ operation (green).
    The blue dots corresponds to the square root of purity obtained by processing the interleaved XEB data with the speckle purity benchmarking technique. The dashed line is an exponential fit to the square root of purity.
    (c) The CZ gate quantum process tomography. The reconstructed Pauli transfer matrix gives the fidelity of $83.3\%$.}
    \label{fig:bench}
\end{figure*}
Withal, the second \ilyas{error minimum}, marked with a star, shows the gate fidelity of 0.9997 with 43.3-ns duration. 
For this point, a time evolution of the mean target state population (dashed blue), a conditional phase (red) and a leakage error (dashed-dotted green) are depicted in Fig.~\ref{fig:pulse}(c). 
The target state population behavior mostly relates to the oscillations between states during the gate implementation. 
The population leaks to the second excited qubit states $|gf\rangle, |fg\rangle$ and to the coupler state $|010\rangle$.
The detailed dynamic of each computational state can be found in Appendix~\ref{sec:Appendix2}.

\subsection{\label{sec:implementation}CZ gate implementation}

We implemented the proposed CZ gate with a total pulse duration of 60 ns\ilyas{, including 9-ns edge segments,} at an amplitude of $-0.28$~V. Since at this amplitude the experimental ZZ interaction strength is \ilyas{slightly larger than in the model, the corresponding gate evolution in the experiment is shorter}. Therefore, we assume that the performed gate corresponds to a point near the third minimum of the error function in Fig.~\ref{fig:pulse}(a).
In order to validate the two-qubit gate performance we start with the quantum process tomography experiment, the result of which is depicted in Fig.~\ref{fig:bench}(c). The fidelity of the reconstructed CZ Pauli transfer matrix is $83.3\%$. 
\lena{We attribute the low fidelity to measurement errors, specifically low-fidelity multi-qubit readout.}
Therefore, we utilize cross-entropy benchmarking (XEB) \cite{Arute2019} resilient against SPAM errors in order to characterize the performance of the implemented CZ operation.

For the XEB we execute the gate sequence shown in Fig.~\ref{fig:bench}(a), where $C_{ij}$ are randomly chosen single-qubit Clifford gates. Here the first index corresponds to the qubit number and the second index refers to the serial number of the gate in the sequence. 
The single-qubit Clifford group is constructed of the $\pi/2$ rotations around X axis of $46.66$ ns duration and virtual S gates in the same way as it was performed in Ref. \cite{Simakov_2023}.
First, we accomplish reference XEB consisting \lena{of} only single-qubit operations simultaneously \lena{executed} on both qubits. 
The obtained depolarization fidelity averaged over randomly sampled quantum circuits as a function of a circuit depth $m$ is shown in Fig.~\ref{fig:bench}(b) with a blue color. 
We approximate the data with the function $a p^m$, where $p$ is the depolarizing parameter and $a$ is a fitting parameter used to take into account a state preparation and measurement (SPAM) errors.
The resulted depolarization parameter $p_1$ is $0.9714 \pm 0.0005$.
The similar data collected for the same circuits with the interleaved CZ gate is depicted in Fig.~\ref{fig:bench}(b) in green. 
The depolarization parameter $p_2$ is equal to $0.947 \pm 0.001$. In order to calculate the conventional fidelity $F$ of a target two-qubit gate we use the formula \cite{Arute2019, Ficheux2021}
\begin{equation}
    F = p + (1-p)/D,
    \label{eq:Fdep}
\end{equation}
where $p=p_2/p_1$. The obtained CZ gate fidelity $F_\mathrm{CZ}$ is $(98.14 \pm 0.12) \%$.
A detailed analysis of errors, along with the corresponding calculations, is provided in Appendix~\ref{sec:Appendix_error_analisys}. Below, we outline the key points.

The total error in interleaved benchmarking arises from single-qubit gate errors and two-qubit CZ gate errors. Single-qubit errors primarily caused by relaxation, dephasing, and static ZZ interaction processes. Based on measured relaxation and coherence times (Table~\ref{tab:gate params}), we estimate a depolarization error of 0.015 due to amplitude and phase damping, while the contribution from static ZZ interaction adds 0.012. The combined effect yields an overall single-qubit gate fidelity of approximately 98.0\%, which agrees well with experiment.  

For the CZ gate, error analysis is more complex due to the interplay of relaxation, dephasing, conditional phase error, leakage, and state population exchange. We begin by estimating the error arising from qubit relaxation and dephasing. In preliminary measurements, we observe that the coherence time \(T_2^*\) decreases by approximately a factor of two when the coupler is tuned to the gate activation flux point, while the relaxation time remains stable across the operating flux range. Assuming that during gate operation \(T_2^*\) is reduced by half compared to its value in the idle state, we estimate a depolarization error of 0.016.
The phase error $\delta\theta = 0.03 \pi$, estimated by XEB method, contributes an additional error of less than 0.002. Leakage out of the computational subspace is assessed by extracting state purity from XEB data using the speckle purity benchmarking technique, yielding a leakage error of 0.008.

Taken together, these results highlight decoherence processes as the dominant error source and particularly underscore the need for suppressing the static ZZ interaction.

\begin{table}
    \centering
    \begin{tabularx}{\columnwidth}{*7{>{\centering\arraybackslash}X}@{}}
    \hline
    \hline
            Qubit & $f_q$ & $\delta_q$ & $T_2^*$ & $T_1$ & $t_\text{1Q}$ & $F_\text{1Q}$ \\
            & (GHz) & (MHz) & $(\mu s)$ & $(\mu s)$ & (ns) & (\%) \\
    \hline
           A & 6.450 & -156 & 8.6 & 15.8 & $2\cdot 46.66$ & 99.1   \\
        B & 6.494 & -157 & 10.8 & 8.3 & $2\cdot 46.66$ & 98.3  \\ 
    \hline
    \hline
    \end{tabularx}
    \caption{The device parameters: qubits frequency, anharmonicity, dephasing and coherence times, and the length of single-qubit Clifford gates in Fig.~\ref{fig:bench}(a). The coefficient 2 denotes the fact, that we construct the Clifford group out of two $R_\mathrm{X}(\pi/2)$ gates with duration of 46.66-ns. \ilyas{The single-qubit gate fidelities are obtained by cross-entropy benchmarking performed simultaneously on both qubits.}}
    \label{tab:gate params}
\end{table}

\section{\label{sec:conclusion}Conclusion}

To conclude, we proposed and demonstrated here a key element for a high-performance scalable architecture based on concentric weakly flux-tunable transmon qubits. In the implemented scheme, the qubits are coupled via the tunable three-mode coupler, which is formed by the extended coplanar capacitor with the Josephson junction at each end and the asymmetric SQUID in the middle. 
We provide the comprehensive numerical analysis of the system and evaluate its characteristics. We show good agreement between the measured ZZ interaction strength and the evaluated from the model with parameters extracted from the experiment. 
Finally, in our architecture we implement the native CZ gate with the pulse duration of 60~ns achieving the two-qubit gate fidelity above 98$\%$, limited mostly by qubit coherence time. The gate fidelity can be further increased by optimizing the fabrication processes and qubit control line parameters. We performed the CZ gate simulation with experimental device parameters that predicts a gate error less than $10^{-3}$ for the optimal gate pulse duration of about 43 ns.

\section*{\label{sec:spasibo}Acknowledgments}

The authors acknowledge Alexey Ustinov for fruitful discussions and useful comments on the manuscript.
The work was supported by Rosatom in the framework of the Roadmap for Quantum computing (Contract No. 151/21-503, December 21, 2021 dated 12/21/2021).
Theoretical analysis is partially supported by the Priority 2030 program at the National University of Science and Technology “MISIS” (Strategic Project Quantum Internet).
\color{black}
The sample was fabricated using the equipment of MIPT Shared Facilities Center (Contract No.868/221-D dated October 24, 2022).

\appendix
\section{\label{sec:Appendix1}Full-circuit Hamiltonian}

\begin{figure*}
    \includegraphics[width=0.8\textwidth]{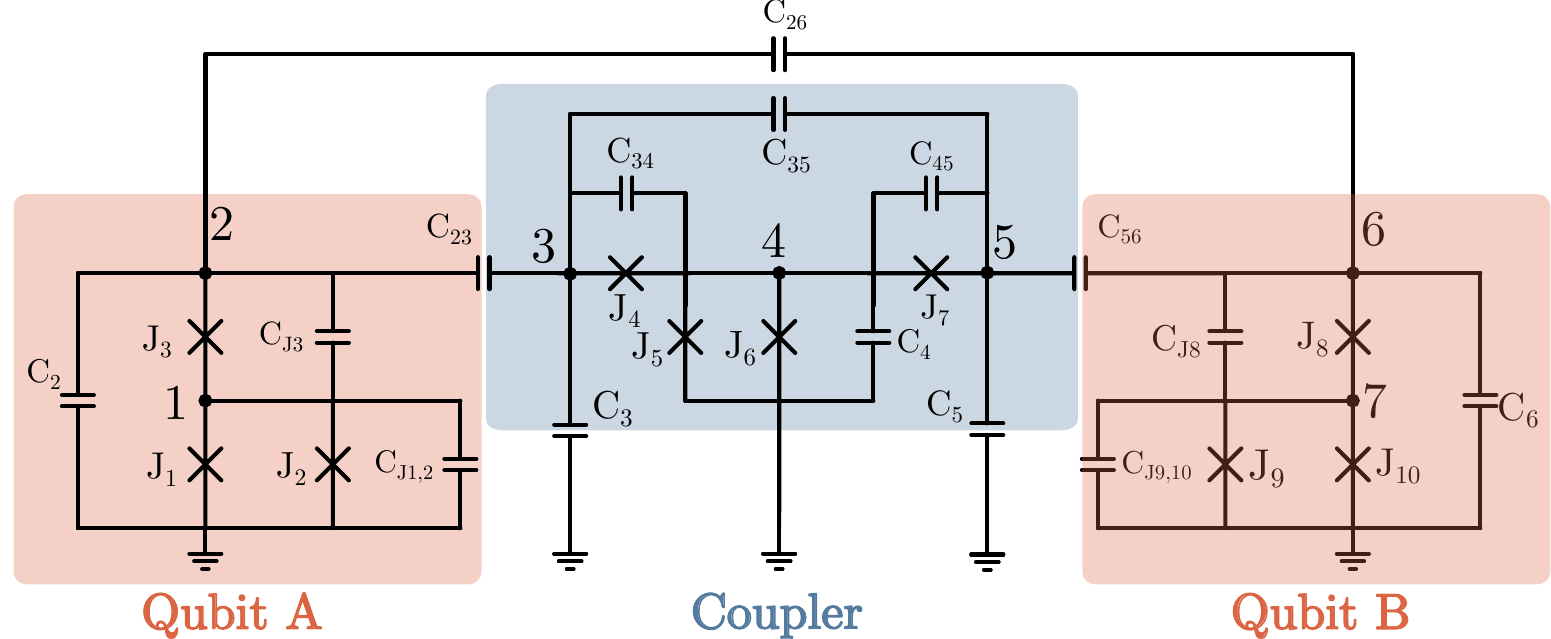} 
	\caption{Equivalent lumped-element circuit for the proposed two-qubit system with a tunable coupler. Transmons (red) are connected through mutual capacitance with each other and with the coupler (blue). $C_i$ stands for the node capacitances with respect to the ground node, $C_{J_i}$ ($i = 1,\dots,10$) are the capacitances of Josephson junctions, $C_{J1,2} = C_{J1}+C_{J2}$ $C_{J19,10} = C_{J9}+C_{J10}$ are total capacitances of SQUID junctions of two transmons, $C_{ij}$ are the mutual capacitances between nodes $i$ and $j$ that facilitate coupling between qubits. $J_{i}$ stands for Josephson inductance of the $i$-th junction. In order not to pile up the scheme, the coupler Josephson junction capacitances are not shown separately, but included in the capacitances between the corresponding nodes.}
    \label{fig:electrical scheme}
\end{figure*}

The corresponding electrical circuit of the proposed two-qubit design with the three-mode coupler is shown in~Fig.~\ref{fig:electrical scheme}. We identified ten Josephson junctions $J_i$, where $i=1,\dots,10$, shunted by capacitances $C_i$ with $i=2,3,4,5,6$. The Josephson junction capacitances are denoted as $C_{Ji}$, where $i=1,2,3,8,9,10$.  The coupler Josephson junction capacitances $C_{J4}$, $C_{J5}$, $C_{J6}$ and $C_{J7}$ are taken into account in corresponding total capacitances $C_{34}$, $C_{4}$, $C_{45}$. The mutual capacitances between the circuit nodes are indicated as $C_{ij}$, where $i$ and $j$ are the node numbers. The external fluxes can be applied to each SQUID of both qubits and the coupler.

The circuit parameter values used for the design are given in Table~\ref{tab:params}. This choice of parameters allows reaching the optimal range of an effective qubit-qubit interaction strength to have the interaction, tending to zero, and enable it to implement a fast two-qubit operation. All junctions are supposed to be fabricated by the shadow evaporation technique with critical current density $j=1~\mu \textnormal{A}/\mu \textnormal{m}^2$.

\begin{table*}
    \centering
    \begin{tabularx}{\textwidth}{ 
     >{\centering\arraybackslash}X 
     >{\centering\arraybackslash}X 
     >{\centering\arraybackslash}X| 
     >{\centering\arraybackslash}X
     >{\centering\arraybackslash}X 
     >{\centering\arraybackslash}X| 
     >{\centering\arraybackslash}X 
     >{\centering\arraybackslash}X }
    \hline
    \hline
    \multicolumn{3}{c|}{Josephson Energy (GHz)} & \multicolumn{3}{c|}{Josephson Capacitance (fF)} & \multicolumn{2}{c}{Node Capacitance (fF)} \\
    \hline
     & Design & Experiment &  & Design & Experiment &  & \\
    \hline
    $E_{J1}$  & 124.2 & 104.4 & $C_{J1}$  & 11.1 & 9.3  & $C_{2}$  & 75.6 \\
    $E_{J2}$  & 43.5  & 23.1  & $C_{J2}$  & 3.9  & 2.1  & $C_{3}$  & 47.2 \\
    $E_{J3}$  & 17.4  & 28.8  & $C_{J3}$  & 1.5  & 2.6  & $C_{4}$  & 144 \\
    $E_{J4}$  & 124.2 & 159.8 & $C_{J4}$  & 11.1 & 14.2  & $C_{5}$  & 46.5 \\
    $E_{J5}$  & 79.5  & 191.5 & $C_{J5}$  & 11.5 & 17.1  & $C_{6}$  & 75.9 \\
    $E_{J6}$  & 129.1 & 117.0 & $C_{J6}$  & 7.1  & 10.4  & $C_{23}$ & 9 \\
    $E_{J7}$  & 124.2 & 159.8 & $C_{J7}$  & 11.1 & 14.2  & $C_{26}$ & 0.015 \\
    $E_{J8}$  & 17.4  & 29.3  & $C_{J8}$  & 1.5  & 2.6  & $C_{34}$ & 13 \\
    $E_{J9}$  & 43.5  & 23.5  & $C_{J9}$  & 3.9  & 2.1  & $C_{35}$ & 0.005 \\
    $E_{J10}$ & 124.2 & 106.1 & $C_{J10}$ & 11.1 & 9.5  & $C_{56}$ & 9.5 \\
    \hline
    \hline
    \end{tabularx}
    \caption{Target circuit parameters of the device: Josephson energies of junctions, node and Josephson junction capacitances. The design parameters of Josephson junctions are compared with the values extracted from the experiment. The coupler Josephson junction capacitances $C_{J4}$, $C_{J5}$, $C_{J6}$, and $C_{J7}$ are included in corresponding values $C_{34}$, $C_{4}$, and $C_{45}$.}
    \label{tab:params}
\end{table*}

Below, we describe each element of the device and derive the full system Hamiltonian.

The single-qubit Hamiltonian has the following form \cite{egorova2024}:
\begin{equation}
    \begin{split}
        \hat{H}_{\text{q}} & =4E_{\textnormal{C}_q}\hat{n}^2+E_{\textnormal{J3}} \left(1-\cos \hat{\varphi}_2\right) +\\
        & +(E_{\textnormal{J1}}+E_{\textnormal{J2}}) \cos \frac{\pi \Phi}{\Phi_0}\sqrt{1+d^2 \operatorname{tan}^2 \frac{\pi \Phi}{\Phi_0}}\left(1-\cos \hat{\varphi}_1\right), 
    \end{split}
\end{equation}
where $\hat{\varphi}_1$, $\hat{\varphi}_2$ are the phase operators on junctions $J_1$, $J_3$ correspondingly, $\hat{n}$ is the Cooper pair number operator canonically conjugated to the sum $\hat{\varphi}=\hat{\varphi}_1+\hat{\varphi}_2$. 
The charge energy is expressed as $E_{\textnormal{C}_q}=e^2/2C_\Sigma$, where $C_\Sigma = C_2+C_{J3}C_{J1,2}/(C_{J3}+C_{J1,2})$. $E_{Ji}$ is Josephson energy of the $i$-th junction, $d=(E_{\textnormal{J2}}-E_{\textnormal{J1}})/(E_{\textnormal{J2}}+E_{\textnormal{J1}})$ is the SQUID asymmetry.

The coupler has three nodes, and therefore we obtain three coupler modes. Node phases can be conveniently expressed with the normal mode coordinates $\widetilde{\varphi}_3, \widetilde{\varphi}_4, \widetilde{\varphi}_5$:
\begin{equation}
\varphi_3 = \widetilde{\varphi}_3 + \widetilde{\varphi}_4;\quad 
 \varphi_4 = \widetilde{\varphi}_3;\quad
 \varphi_5 = \widetilde{\varphi}_3 - \widetilde{\varphi}_5. 
\label{eq1}     
\end{equation}
Here and further, we omit the hats of quantum operators in order to avoid an overloading of the formulas.
Hence, we have the coupler Hamiltonian
\begin{widetext}
\begin{equation}
    \begin{split}
        H_\text{c} &= 4E_{C_{00}}\widetilde{n}_3^2 + 4E_{C_{11}}\widetilde{n}_4^2 + 4E_{C_{22}}\widetilde{n}_5^2 +  2E_{C_{01}}\widetilde{n}_3 \widetilde{n}_4 + 2E_{C_{02}}\widetilde{n}_3 \widetilde{n}_5 + 
2E_{C_{12}}\widetilde{n}_4 \widetilde{n}_5 +\\
&+ E_{J4}\left(1-\cos \widetilde{\varphi}_4\right) + E_{J_7}\left(1-\cos \widetilde{\varphi}_5\right) + (E_\textnormal{J5} +E_\textnormal{J6}) \cos \frac{\pi \Phi}{\Phi_0}\sqrt{1+d_{}^2 \textnormal{tan}^2 \frac{\pi \Phi}{\Phi_0}}(1-\cos \widetilde{\varphi}_3).
\label{coup_ham}
    \end{split}
\end{equation}

Here $E_{C_{ij}}=e^2/2C_{ij}$, where $C_{ij}$ is the element of a $3\times3$ capacitance matrix:
\begin{equation}
\frac{1}{2}\begin{bmatrix}
C_\textnormal{3}+C_\textnormal{4}+C_\textnormal{5}&C_\textnormal{3}&-C_\textnormal{5}\\
C_\textnormal{3}&C_\textnormal{3}+C_\textnormal{34}+C_\textnormal{35}&C_\textnormal{35}\\
-C_\textnormal{5}&C_\textnormal{35}&C_\textnormal{35}+C_\textnormal{45}+C_\textnormal{5}
\end{bmatrix}.
\label{eqA4}
\end{equation}

Eventually, the coupler Hamiltonian can be represented in the shortened form as three subsystems with capacitive interaction between each other:
\begin{equation}
    H_\text{c} = \sum_{i}^{} H_{i} +\sum_{i\not= j}^{} H_{i j},\quad
i,j  \in \{3, 4, 5\}.
\label{eqA13}
\end{equation}
Here, the index $i$ corresponds to the Hamiltonian terms associated with the $i$-th phase coordinate.

For the proposed system we choose node fluxes $\varphi_i$, corresponding to nodes $i$ in Fig.~\ref{fig:electrical scheme}, as the generalized coordinates of the system. We can write down the circuit Lagrangian $L(\varphi_i,\dot{\varphi_i})$ using node fluxes together with the voltages $\dot{\varphi}_i$:

\begin{equation}
   L = T - U,  
\end{equation}
\begin{equation}
    \begin{split}
        T = \frac{\Phi^2_0}{2(2\pi)^2}\Big[&C_\textnormal{J1,2}\dot{\varphi}_1^2 + C_2\dot{\varphi}_2^2 + C_\textnormal{J3}(\dot{\varphi}_2-\dot{\varphi}_1)^2 + C_\textnormal{23}(\dot{\varphi}_3-\dot{\varphi}_2)^2 +  C_3\dot{\varphi}_3^2 + C_\textnormal{34}(\dot{\varphi}_4-\dot{\varphi}_3)^2 + \\
        & +C_4\dot{\varphi}_4^2 + C_5\dot{\varphi}_5^2 + C_\textnormal{45}(\dot{\varphi}_5-\dot{\varphi}_4)^2 +C_\textnormal{35}(\dot{\varphi}_5-\dot{\varphi}_3)^2 + C_6\dot{\varphi}_6^2 + \\ & + C_\textnormal{56}(\dot{\varphi}_6-\dot{\varphi}_5)^2 + C_\textnormal{26}(\dot{\varphi}_6-\dot{\varphi}_2)^2 + C_\textnormal{J8}(\dot{\varphi}_7-\dot{\varphi}_6)^2 + C_\textnormal{J9,10}\dot{\varphi}_7^2 \Big],
    \end{split}
    \label{eqA2}
\end{equation}
\begin{equation}
    \begin{split}
        U &= E_\textnormal{J1}\big(1-\cos\varphi_1\big) + E_\textnormal{J2}\big(1-\cos(\varphi_1-\varphi_{\text{ext}, 1})\big) + E_\textnormal{J3}\big(1-\cos(\varphi_2-\varphi_1)\big) + \\ 
        & + E_\textnormal{J4}\big(1-\cos(\varphi_4-\varphi_3)\big) + E_\textnormal{J5}\big(1-\cos\varphi_4\big) + E_\textnormal{J6}\big(1-\cos(\varphi_4-\varphi_{\text{ext}, c})\big) + \\
        & + E_\textnormal{J7}\big(1-\cos(\varphi_5-\varphi_4)\big) + E_\textnormal{J8}\big(1-\cos(\varphi_6-\varphi_7)\big) + E_\textnormal{J9}\big(1-\cos\varphi_7\big)+ \\ & + E_\textnormal{J10}\big(1-\cos(\varphi_7-\varphi_{\text{ext}, 2})\big),
    \end{split}
    \label{eqA3}
\end{equation}
where $T$ and $U$ are, respectively, the kinetic and potential energy. 
The kinetic energy term can be rewritten in matrix form $T=\frac{1}{2}\vec{\dot{\varphi}}^T C \vec{\dot{\varphi}}$, where $\vec{\dot{\varphi}}=[\dot{\varphi}_1, \dot{\varphi}_2, \dot{\varphi}_3, \dot{\varphi}_4, \dot{\varphi}_5, \dot{\varphi}_6, \dot{\varphi}_7]$ and $C$ is a $7\times7$ capacitance matrix:
\begin{equation}
C=\begin{bmatrix}
C_\textnormal{1$\Sigma$}&-C_{\textnormal{J3}}&0&0&0&0&0\\
-C_{\textnormal{J3}}&C_\textnormal{2$\Sigma$}&-C_{23}&0&0&-C_{26}&0\\
0&-C_{23}&C_\textnormal{3$\Sigma$}&-C_{\textnormal{34}}&-C_{35}&0&0\\
0&0&-C_{\textnormal{34}}&C_\textnormal{4$\Sigma$}&-C_{45}&0&0\\
0&0&-C_{35}&-C_{45}&C_\textnormal{5$\Sigma$}&-C_{\textnormal{56}}&0\\
0&-C_{26}&0&0&-C_{56}&C_\textnormal{6$\Sigma$}&-C_{\textnormal{J8}}\\
0&0&0&0&0&-C_\textnormal{J8}&C_{\textnormal{7$\Sigma$}}\\
\end{bmatrix}, \\
\label{eqA4}
\end{equation}
where
\begin{equation}
    \begin{split}
        C_\textnormal{1$\Sigma$} &= C_\textnormal{J1,2}+C_\textnormal{J3}, \\
        C_\textnormal{2$\Sigma$} &= C_\textnormal{2}+C_\textnormal{J3}+C_\textnormal{23}+C_\textnormal{26}, \\
        C_\textnormal{3$\Sigma$} &= C_\textnormal{3}+C_\textnormal{23}+C_\textnormal{34}+C_\textnormal{35}, \\
        C_\textnormal{4$\Sigma$} &= C_\textnormal{4}+C_\textnormal{34}+C_\textnormal{45}, \\
        C_\textnormal{5$\Sigma$} &= C_\textnormal{5}+C_\textnormal{45}+C_\textnormal{35}+C_\textnormal{56}, \\
        C_\textnormal{6$\Sigma$} &= C_\textnormal{6}+C_\textnormal{56}+C_\textnormal{26}+C_\textnormal{J8}, \\
        C_\textnormal{7$\Sigma$} &= C_\textnormal{J8}+C_\textnormal{J9,10}.
    \end{split}
    \label{eqA5}
\end{equation}

Applying the coupler coordinate transformation (Eq.~\ref{eq1}) to the capacitance matrix yields
\begin{equation}
C_\textnormal{new}=T_r^T \times C \times T_r,\quad  
\end{equation}
where the transformation matrix $T_r$ is defined as:
\begin{equation}
T_r=\frac{1}{2}\begin{bmatrix}
1&0&0&0&0&0&0\\
0&1&0&0&0&0&0\\
0&0&1&1&0&0&0\\
0&0&1&0&0&0&0\\
0&0&1&0&-1&0&0\\
0&0&0&0&0&1&0\\
0&0&0&0&0&0&1\\
\end{bmatrix}. \\
\label{eqA6}
\end{equation}

The potential energy becomes

\begin{equation}
    \begin{split}
        U &= E_\textnormal{J1}\big(1-\cos\varphi_1\big) + E_\textnormal{J2}\big(1-\cos(\varphi_1-\varphi_{\text{ext}, 1})\big) + E_\textnormal{J3}\big(1-\cos(\varphi_2-\varphi_1)\big) + \\ 
        & + E_\textnormal{J4}\big(1-\cos\widetilde{\varphi}_4\big) + E_\textnormal{J5}\big(1-\cos\widetilde{\varphi}_3\big) + E_\textnormal{J6}\big(1-\cos(\widetilde{\varphi}_3-\varphi_{\text{ext}, c})\big) + E_\textnormal{J7}\big(1-\cos\widetilde{\varphi}_5\big) + \\
        & + E_\textnormal{J8}\big(1-\cos(\varphi_6-\varphi_7)\big) + E_\textnormal{J9}\big(1-\cos\varphi_7\big) + E_\textnormal{J10}\big(1-\cos(\varphi_7-\varphi_{\text{ext}, 2})\big).
    \end{split}
    \label{eqA7}
\end{equation}
\end{widetext}

\begin{figure}
    \centering
    \includegraphics[width=\columnwidth]{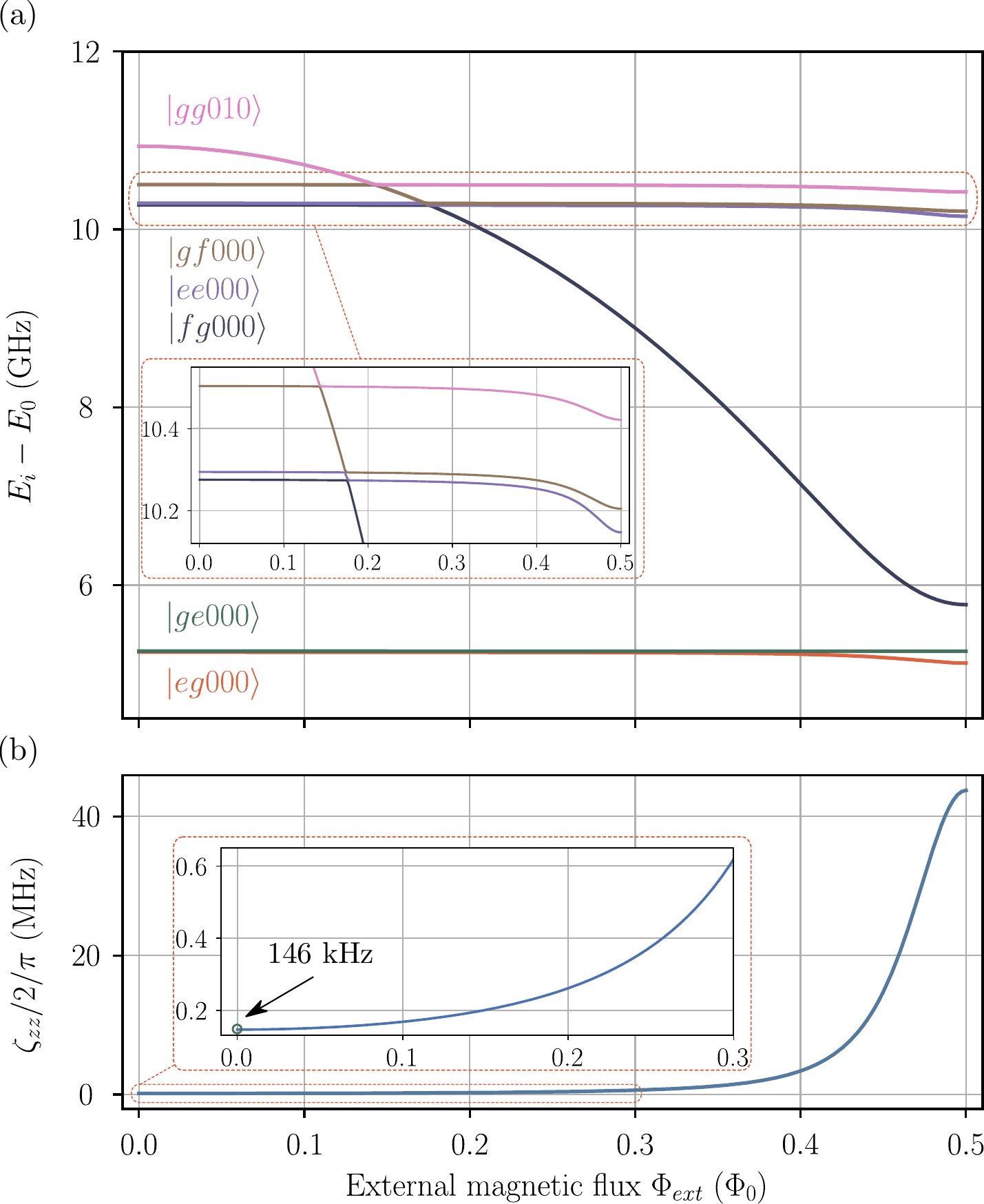}
\caption{(a) Energy levels of the transmon-coupler-transmon system from external magnetic flux in the coupler with parameters from Tab.II. The system is determined by the states of five subsystems. The first symbol in the system quantum state corresponds to the state of the first transmon ($|g\rangle$, $|e\rangle$ or $|f\rangle$), the second symbol corresponds to the state of the second transmon (similarly to the first transmon), the rest corresponds to the state of the coupler. Coupler tunable mode pushes transmon levels, changing the value of ZZ interaction between qubits. The system states that are important when considering population dynamics (see Fig.~\ref{fig:populations}) are noted.
(b) ZZ interaction between qubits vs external magnetic flux on the coupler. The inset shows the coupling near the zero flux bias point, illustrating the rate of its increase. The green circle shows the minimal value of ZZ interaction.}
\label{fig:spectrum}
\end{figure}
By means of this transformation the coupler circuit can be divided into three subsystems. Thus, the matrices for calculation have a smaller dimension and calculations are faster. Taking into account two transmons, we have five subsystems with five levels in each one. We define grid for each phase $\varphi_i$ variable with phase step $\delta\varphi_i = \pi/16$ and corresponding canonically conjugate momenta $q_i = \frac{\partial  L}{\partial {\dot{\varphi}}_i}.$ The grid for coupler external flux has the phase step $\delta\varphi_{ext} = \pi/128.$
The system Hamiltonian in terms of the first-order normal modes is defined as
\begin{equation}
H = \sum_{i}^{} q_i {\dot{\varphi}}_i - L =\frac{1}{2} \vec{q}^T C_{\text{new}}^{-1} \vec{q} + U, 
\label{eqA8}
\end{equation}
where $C_{\text{new}}^{-1}$ is the inverse capacitance matrix.
We calculate the eigenvalues of each subsystem in phase basis using numerical diagonalization and find charge matrix elements by means of Fourier transformation. Then we construct the full Hamiltonian 
\lena{
\begin{equation}
\hat{H}_\mathrm{sys} = \sum_{i \in \{ {Q_\mathrm{A}, Q_\mathrm{B}, c_1, c_2, c_3 \}}} \hat{H}_i + \hat{H}_\text{int},
\label{full_ham}
\end{equation}
where we consider our system as five subsystems associated with 5 nodes in Fig.~\ref{fig:qcq_zoom}(c). 
Two of nodes $Q_\mathrm{A}$ and $Q_\mathrm{B}$ correspond to the qubits, while the rest three $c_1, c_2, c_3$ refer to the coupler modes. 
The interaction term takes the following form:
\begin{equation}
      \hat{H}_\text{int} = \sum_{i \neq j}E_{Cij}\hat{n}_i \hat{n}_j,  
      \label{int_ham}
\end{equation}
where $\{i,j\}$ is the set of circuit nodes pairs, \ilyas{$\hat{n}$ is the Cooper pair number operator} and $E_{Cij}$ is the energy of the capacitive interaction. }
Charge energy here is defined as $E_{Cij} = \frac{e^2}{2}C_{\text{new}}^{-1}.$

Finally, we numerically diagonalized the full Hamiltonian. 
The calculated energy spectrum as a function of magnetic flux $\Phi$ in the coupler SQUID is plotted in Fig.~\ref{fig:spectrum}(a). 
We provide the full system spectrum calculated for the qubits at the operating point as a function of an external magnetic flux $\Phi_\text{ext}$. The eigenstates are labelled as $|n_\text{q1},n_\text{q2}, n_\text{c1}, n_\text{c2}, n_\text{c3}\rangle$, where $n_{i}$ is the occupancy of the $i$-th mode. The four computational levels are $|gg000\rangle$, $|eg000\rangle$, $|ge000\rangle$, $|ee000\rangle$.

As our goal is to implement a CZ gate, we are interested in the controllable ZZ interaction between the computational qubits. Here, the effective ZZ interaction is determined according to the conventional expression $\zeta_{ZZ} = f_{ee} - f_{eg} - f_{ge}$. We numerically calculate the strength of the interaction for the transmon qubits, placed at upper sweet spots, and depict the result in Fig.~\ref{fig:spectrum}(b). It can be seen that varying the external magnetic flux on the coupling element, one can change the effective interaction from 146 kHz up to 44 MHz. Thereby, the ZZ coupling is tuned off at the zero flux bias point ($\Phi= 0$) and activated at the flux degeneracy point ($\Phi= 0.5 \Phi_0$).

\begin{widetext}

\section{\label{sec:Appendix_ZZ}Derivation of the XX and ZZ couplings}
In the linear approximation, the Josephson junction capacitances vanish from the electrical scheme in Fig.~\ref{fig:electrical scheme}, and qubits A and B are associated with nodes 2 and 6, respectively. The capacitances $C_{34}$, $C_{45}$ and $C_{35}$ are negligible and therefore not taken into account. The direct coupling $C_{26}$ will be included later. The capacitance and inverse inductive matrices, expressed in terms of node phases $\varphi_i$, are given by: 
\begin{equation}
\tilde{C}=\begin{bmatrix}
C_\textnormal{23}+C_{\mathrm{q_A}}&-C_\textnormal{23}&0&0&0\\
-C_{\textnormal{23}}&C_\textnormal{23}+C_{3}&0&0&0\\
0&0&C_\textnormal{4}&0&0\\
0&0&0&C_\textnormal{5}+C_{56}&-C_{56}\\
0&0&0&-C_{\textnormal{56}}&C_\textnormal{56}+C_{\mathrm{q_B}}\\
\end{bmatrix}, \\
\end{equation}

\begin{equation}
\tilde{L}^{-1}=\begin{bmatrix}
0&0&0&0&0\\
0&\frac{1}{L_\mathrm{J4}}&-\frac{1}{L_\mathrm{J4}}&0&0\\
0&-\frac{1}{L_\mathrm{J4}}&\frac{1}{L_\mathrm{J4}}+\frac{1}{L_\mathrm{J56}}+\frac{1}{L_\mathrm{J7}}&-\frac{1}{L_\mathrm{J7}}&0\\
0&0&-\frac{1}{L_\mathrm{J7}}&\frac{1}{L_\mathrm{J7}}&0\\
0&0&0&0&0\\
\end{bmatrix}. \\
\end{equation}

After applying the transformation
\begin{equation}
T_\text{diag}=\begin{bmatrix}
1&0&0&0&0\\
0&1&1&0&0\\
0&0&1&0&0\\
0&0&1&1&0\\
0&0&1&0&1\\
\end{bmatrix}. \\
\end{equation}

the inductance matrix takes the diagonal form:
\begin{equation}
L^{-1}=\begin{bmatrix}
0&0&0&0&0\\
0&\frac{1}{L_\mathrm{J4}}&0&0&0\\
0&0&\frac{1}{L_\mathrm{J56}}&0&0\\
0&0&0&\frac{1}{L_\mathrm{J7}}&0\\
0&0&0&0&0\\
\end{bmatrix}. \\
\end{equation}

The Lagrange function, in the presence of an external current $I_\mathrm{q_A}$ from qubit A, is expressed as:
\begin{equation}
      L(\vec{\Phi},\dot{\vec{\Phi}}) = \frac{1}{2}\dot{\vec{\Phi}}^T C \dot{\vec{\Phi}} - \frac{1}{2} \vec{\Phi}^T L^{-1} \vec{\Phi} + I_\mathrm{q_A}(t)\Phi_2.  
\end{equation}

Using the Euler-Lagrange equation, a system of equations of motion is derived:
\begin{equation}
      C\ddot{\vec{\Phi}} + L^{-1}\vec{\Phi} = I_\mathrm{q_A}(t)\vec{e},
\end{equation}
where $\vec{e}^T = (1,0,0,0,0)$. Substituting the time-dependent flux 
$\vec{\widetilde{\Phi}}(t) = \vec{\tilde{\Phi}}e^{i\omega t}$ into the system of motion equations yields the expression describing the relationship between node fluxes and the induced current:
\begin{equation}
      (-\omega^2C + L^{-1})\vec{\widetilde{\Phi}} = \tilde{I}_\mathrm{q_A}\vec{e}.
      \label{ind_curr}
\end{equation}

For simplification, we consider a symmetric system in further calculations: 
$C_\textnormal{56} = C_\textnormal{23}$, $C_\textnormal{3} = C_\textnormal{5}$, $C_\mathrm{q_A} = C_\mathrm{q_B}$, $L_\textnormal{J7} = L_\textnormal{J4}.$ Through elementary row transformations, we separate the coupler variables, resulting in the following subsystem equations for the coupler:
\begin{equation}
      (1-\omega^2L_\mathrm{c}C_\mathrm{c})\vec{\widetilde{\Phi}}_\mathrm{c} = \frac{C_{23}}{C_{23}+C_\mathrm{q_A}}\tilde{I}_\mathrm{q_A}\left( \begin{matrix}
L_\textnormal{J4} \\
L_\textnormal{J56} \\
0 \\
\end{matrix}
\right),
\end{equation}
where the coupler inductance and capacitance matrices are:
\begin{equation}
L_\textnormal{c}=\begin{bmatrix}
L_\mathrm{J4}&0&0\\
0&L_\mathrm{J56}&0\\
0&0&L_\mathrm{J4}\\
\end{bmatrix}, \\
\end{equation}

\begin{equation}
C_\textnormal{c}=\begin{bmatrix}
-\frac{C^2_\textnormal{23}}{C_\textnormal{23}+C_\mathrm{q_A}}+C_{23}+C_3&-\frac{C^2_\textnormal{23}}{C_\textnormal{23}+C_\mathrm{q_A}}+C_\textnormal{23}+C_3&0\\
-\frac{C^2_\textnormal{23}}{C_\textnormal{23}+C_\mathrm{q_A}}+C_\textnormal{23}+C_3&-\frac{2C^2_\textnormal{23}}{C_\textnormal{23}+C_\mathrm{q_A}}+2C_\textnormal{23}+2C_3+C_4&-\frac{C^2_\textnormal{23}}{C_\textnormal{23}+C_\mathrm{q_A}}+C_\textnormal{23}+C_3\\
0&-\frac{C^2_\textnormal{23}}{C_\textnormal{23}+C_\mathrm{q_A}}+C_\textnormal{23}+C_3&-\frac{C^2_\textnormal{23}}{C_\textnormal{23}+C_\mathrm{q_A}}+C_\textnormal{23}+C_3\\
\end{bmatrix}. \\
\end{equation}

By defining 
\begin{equation}
    F(\omega)=1-\omega^2L_\textnormal{c}C_\textnormal{c},
    \label{eq:eigencoupler}
\end{equation}
the coupler fluxes are expressed as:
\begin{equation}
\vec{\Phi}_\textnormal{c} = \frac{C_\textnormal{23}}{C_\textnormal{23} + C_\mathrm{q_A}} \tilde{I}_\mathrm{q_A}  \frac{\mathrm{adj}( F(\omega))}{\mathrm{det}(F(\omega))}
\left( \begin{matrix}
L_\textnormal{J4} \\
L_\textnormal{J56} \\
0 \\
\end{matrix}
\right),
\end{equation}

where $\mathrm{det}(F(\omega))=\frac{\omega^2_\mathrm{c_+}\omega^2_\mathrm{c_0}\omega^2_\mathrm{c_\pm}}{(\omega^2_\mathrm{c_+}-\omega^2)(\omega^2_\mathrm{c_0}-\omega^2)(\omega^2_\mathrm{c_-}-\omega^2)}$ is the determinant, which depends on the coupler normal modes $\omega_{c_+}, \omega_{c_0}, \omega_{c_-}$, depicted in Fig.~\ref{fig:qcq_zoom}(d) in ascending order of frequency. The mode indices indicate the relative directions of charge oscillations at the central coupler node with respect to the side nodes.

During the elementary row transformations, the relationship between coupler fluxes becomes apparent: $\omega^2C_\textnormal{23}\tilde{\Phi}_4 + \omega^2C_\textnormal{23}\tilde{\Phi}_5 -\omega^2(C_\mathrm{q_A}+C_\textnormal{23})\tilde{\Phi}_6 =0 $. Together with the expression $\tilde{V}_\mathrm{q_B} = i\omega\tilde{\Phi}_6$, the voltage on the second qubit can be written as $\tilde{V}_\mathrm{q_B} = i\omega\left(\frac{C_\textnormal{23}}{C_\textnormal{23}+C_\mathrm{q_A}}\right)^2\frac{L_\textnormal{J56}\omega^2_\mathrm{c_+}\omega^2_\mathrm{c_0}\omega^2_\mathrm{c_-}}{(\omega^2_\mathrm{c_+}-\omega^2)(\omega^2_\mathrm{c_0}-\omega^2)(\omega^2_\mathrm{c_-}-\omega^2)}\tilde{I}_\mathrm{q_A}$, where the expressions for the determinant and the algebraic complement of the function $F(\omega)$ are substituted.
Consequently, the imaginary part of the impedance between the two qubits, given by $Z_\mathrm{q_A q_B} = \tilde{V}_\mathrm{q_B}/\tilde{I}_\mathrm{q_A}$, is:
\begin{equation}
 \mathrm{Im}Z_\mathrm{q_A q_B}(\omega)= \omega\left(\frac{C_\textnormal{23}}{C_\textnormal{23}+C_\mathrm{q_A}}\right)^2\frac{L_\textnormal{J56}\omega^2_\mathrm{c_+}\omega^2_\mathrm{c_0}\omega^2_\mathrm{c_-}}{(\omega^2_\mathrm{c_+}-\omega^2)(\omega^2_\mathrm{c_0}-\omega^2)(\omega^2_\mathrm{c_-}-\omega^2)}.
\end{equation}

Finally, the XX coupling can be estimated as a function of impedance, as described in \cite{impedance, solgun2022direct}:
\begin{equation}
    J_\mathrm{q_A q_B} = -\frac{1}{4}\sqrt{\frac{\omega_\mathrm{q_A}\omega_\mathrm{q_B}}{L_\mathrm{q_A}L_\mathrm{q_B}}}\textnormal{Im}\left[\frac{Z_\mathrm{q_A q_B}(\omega_\mathrm{q_A})}{\omega_\mathrm{q_A}} + \frac{Z_\mathrm{q_A q_B}(\omega_\mathrm{q_B})}{\omega_\mathrm{q_B}}\right],
\end{equation}
where $L_\mathrm{q_A}$ and $L_\mathrm{q_B}$ are the total inductances of 3-JJ qubit circuits in this case. The total XX coupling between qubits is estimated as $J_{XX} = J_\mathrm{q_A q_B} + J_0$, where $J_0$ represents the direct capacitive interaction due to the presence of $C_{26}$. 

In our system, the ZZ interaction is primarily determined by the shift of the $|ee\rangle$ level due to perturbations within the $|fg\rangle$, $|gf\rangle$, and $|ee\rangle$ manifold. According to the previously mentioned paper \cite{solgun2022direct}, the interaction between the $|fg\rangle$ and $|ee\rangle$ levels, as well as between $|gf\rangle$ and $|ee\rangle$, is approximately $\sqrt{2}J_{XX}$ in the first-order approximation. 

The Hamiltonian of the two-qubit state subspace is constructed using the Duffing oscillator model:
\begin{equation}
        H = \omega_\mathrm{q_A} a ^ \dagger a + \frac{\delta_\mathrm{q_A}}{2} a ^ \dagger a  (a ^ \dagger a  - 1) 
        +\omega_\mathrm{q_B} b ^ \dagger b + \frac{\delta_\mathrm{q_B}}{2} b ^ \dagger b  (b ^ \dagger b  - 1) + J_{XX} (a ^ \dagger b + a b ^ \dagger).
\end{equation}
Here, $J_{XX}$ represents the total XX coupling described above. 
The numerical Hamiltonian is then constructed, which is formed by the energy levels $|fg\rangle$, $|gf\rangle$, and $|ee\rangle$, while accounting for the $\sqrt{2}J_{XX}$ interaction mentioned above. Finally, the ZZ interaction strength is determined as the frequency difference between the $|ee\rangle$ energy level in the undressed and dressed Hamiltonians.

\end{widetext}



\section{\label{sec:AppendixComp} Comparison with a single-mode tunable coupler}

To illustrate the advantages of the implemented three-mode coupler using technologically feasible Josephson junction parameters, we compare the XX interaction strength calculated analytically for two configurations: a conventional single-mode tunable coupler \cite{Yan2018} and the proposed three-mode coupler (see Section~\ref{sec:zz_int}). All system parameters are kept identical, except that in the single-mode case, side junctions are omitted, and the coupler capacitance is taken as the sum of the capacitances of all coplanar parts.

\begin{figure}
    \centering
    \includegraphics[width=\columnwidth]{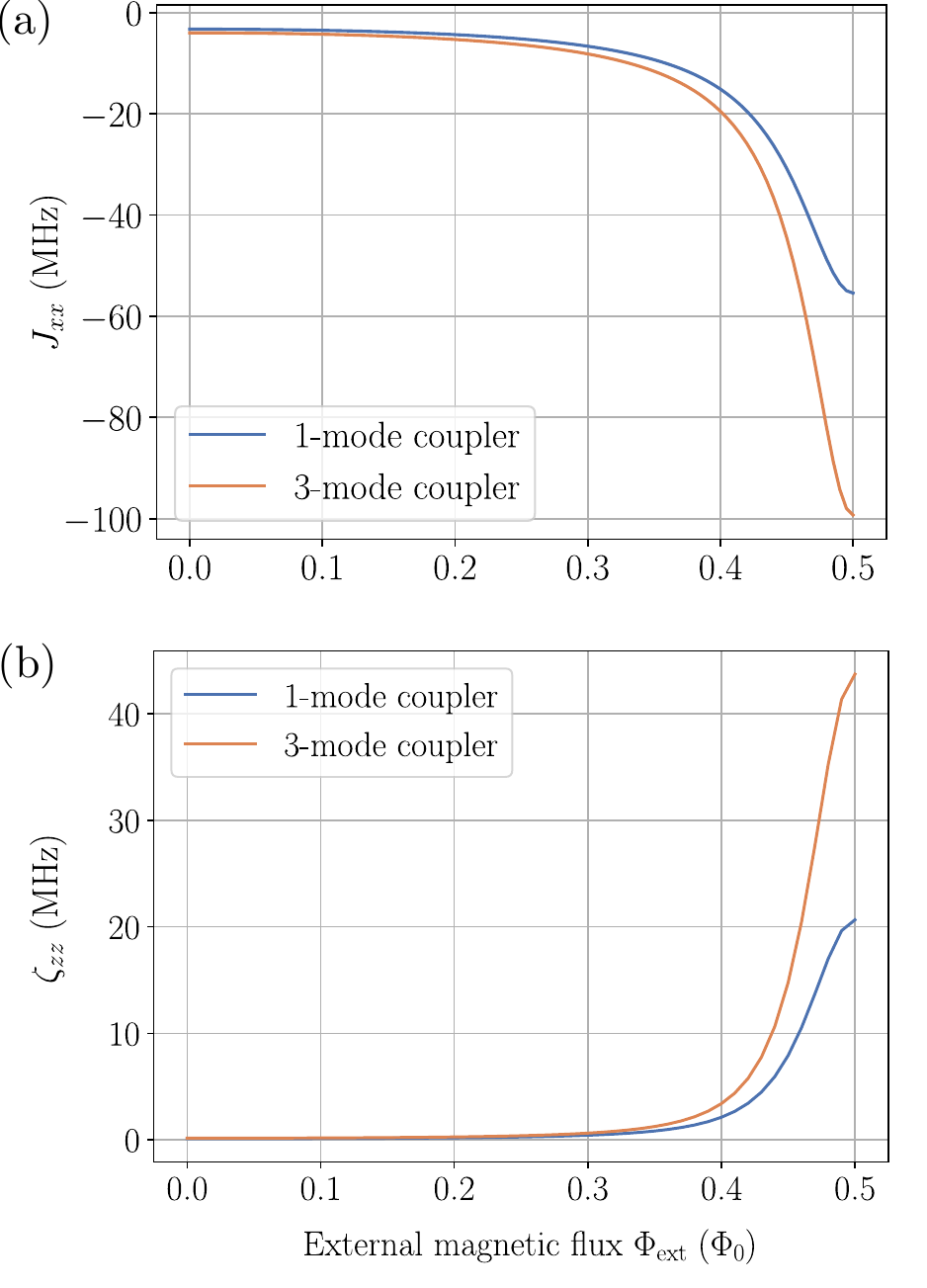}
\caption{Comparison between the single-mode tunable coupler and the three-mode tunable coupler.
(a) XX interaction analytically calculated for both coupler types as a function of external magnetic flux in the coupler SQUID.
(b) ZZ interaction numerically calculated for both coupler types as a function of external magnetic flux in the coupler SQUID.}
\label{fig:coupler comparison}
\end{figure}

As shown in Fig.~\ref{fig:coupler comparison}(a), the tunable range of the XX coupling as a function of magnetic flux in the coupler SQUID is narrower in the single-mode design. Since the ZZ interaction strength is approximately proportional to the square of the XX coupling, the tunable range of the ZZ coupling is also wider in the three-mode configuration, while maintaining residual ZZ interaction strengths comparable to the single-mode one. This conclusion is supported by a numerical simulation, which yield consistent results (Fig.~\ref{fig:coupler comparison}(b)).

\section{\label{sec:AppendixDetuned} ZZ interaction between highly detuned qubits}

To evaluate the coupler performance with highly detuned qubits, we calculated the ZZ interaction strength using a combination of design and experimental parameters (see Tab.~\ref{tab:params}). Specifically, the design parameters were used for the coupler and one qubit, while the Josephson junction areas of the second qubit were set to their experimentally extracted values, resulting in a qubit frequency difference of approximately 1.2 GHz. The calculated ZZ interaction strength varied from 13 kHz to 60 MHz as a function of external magnetic flux. These results demonstrate that, for highly detuned qubits, it is possible to optimize system parameters to achieve both low residual ZZ interaction and sufficient coupling strength at the operating point.

\begin{figure}
    \centering
    \includegraphics[width=\columnwidth]{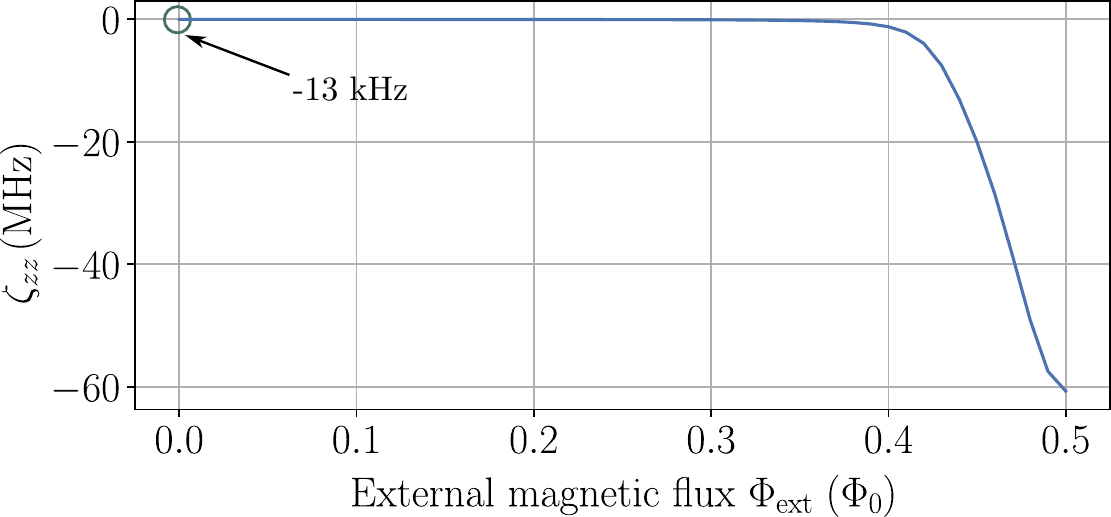}
\caption{Numerically calculated ZZ interaction as a function of external magnetic flux in the coupler SQUID for highly detuned qubits mediated by a three-mode tunable coupler.}
\label{fig:ZZ mixed}
\end{figure}

\color{black}
\section{\label{sec:Appendix2}Time evolution of state population}
\begin{figure*}
    \centering
    \includegraphics[width=\textwidth]{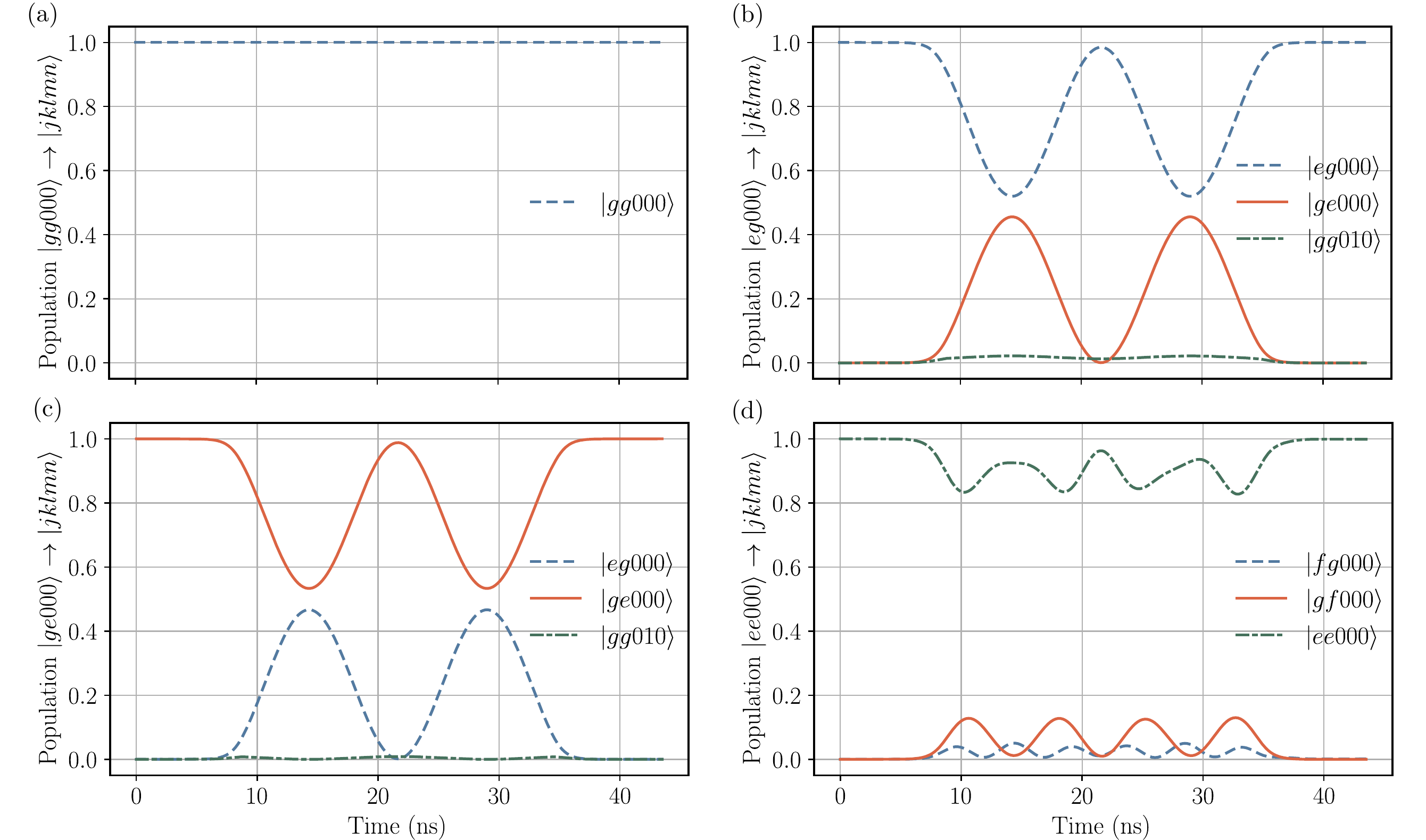}
    \caption{Time evolution of populations for four initial computational states during the implemented gate. The state notation corresponds to the mode occupations of the Hamiltonian. The flux pulse on the coupler has the shape shown in Fig.~\ref{fig:pulse}(d) in the main text. In this case, the fidelity of the CZ gate $F=0.9997$.}
    \label{fig:populations}
\end{figure*}

We investigated the time evolution of the computational states population during the CZ gate (see Fig.~\ref{fig:populations}, corresponds to the star in Fig.~\ref{fig:pulse}(a) in the main text) with experimental device parameters (see Tab.~\ref{tab:params}). Besides exchange between four target states, we have leakages to the levels $|fg000\rangle$, $|gf000\rangle$ and $|gg010\rangle$. At the end of the two-qubit operation, the overall population returns to the computational subspace.

\section{\label{sec:Appendix_error_analisys}Error analysis}

The total error obtained from interleaved benchmarking comprises contributions from single-qubit gate errors and the two-qubit CZ gate error. Let us examine each of these components individually.
We begin with single-qubit errors, which are influenced by qubit dephasing, relaxation, and the static ZZ interaction. 
Using the measured relaxation ($T_1$) and coherence ($T_2^*$) times (Table~\ref{tab:gate params}), we apply a standard operator-sum representation to model these error channels. 
The fidelity of single-qubit gates implemented simultaneously on both qubits is calculated following the formula \cite{li2024realization, PhysRevX.13.031035}
\begin{equation}
    F = 1 - \frac{2t}{5} \sum_{A,B} \left(\frac{1}{T_1} + \frac{1}{T_\varphi}\right),
    \label{eq:fidelity2Q_T1T2}
\end{equation}

where $t$ is the gate duration and the dephasing time $T_\varphi$ is obtained from the relation $\frac{1}{T^*_2} = \frac{1}{2T_1} + \frac{1}{T_\varphi}$.
\color{black}
From this, we compute a fidelity of 98.9\%, which corresponds to a depolarization error of 0.015.
The next contributor to single-qubit errors is the static ZZ interaction when the coupler is inactive. To estimate its impact on fidelity, we use the expression
\begin{equation}
    F = \frac{(7 + 3\cos\varphi)}{10},
\end{equation}
 where $\varphi = 2\pi \zeta_\mathrm{ZZ} t_\text{1Q}$, with $\zeta_\mathrm{ZZ}$ as the ZZ interaction rate with the zero flux at the coupler. Substituting values for $\varphi$, we find a fidelity of 99.1\%, which corresponds to a depolarization error of 0.012. 
Combining these two effects, the overall fidelity of the single-qubit gates is found to be $98.0$\%, which aligns well with the measured fidelity of $97.68 \pm 0.04$\%. This close agreement validates our error model and analysis.

To further substantiate our analysis, we performed single-qubit gate benchmarking for each qubit individually. 
This experiment was conducted using a different experimental setup but with the exact same sample. 
The $R_x(\pi/2)$ gate duration was set to $20~\mathrm{ns}$, with measured relaxation times $T_1 = 16.7$ and $14.5~\mu\mathrm{s}$ and coherence times $T_2^* = 8.9$ and $10.5~\mu\mathrm{s}$ for both qubits A and B. 
The single-qubit fidelities obtained through individual XEB measurements were $99.77\%$ and $99.81\%$ for qubits A and B, respectively. 
Theoretical expectations, calculated using the formula 
\begin{equation}
    F = 1 - \frac{t_\text{1Q}}{3} \left(\frac{1}{T_1} + \frac{1}{T_\varphi}\right), 
\end{equation}
yield values of $99.81\%$ and $99.82\%$. 
These results confirm that, for individual qubits, the fidelity of the gates is predominantly limited by dephasing and relaxation processes.

Next, we analyze the errors associated with the CZ gate, which presents a more complex problem compared to single-qubit gates.
We begin by calculating the error arising from qubit relaxation and dephasing.
\begin{figure*}
    \centering
    \includegraphics[width=\textwidth]{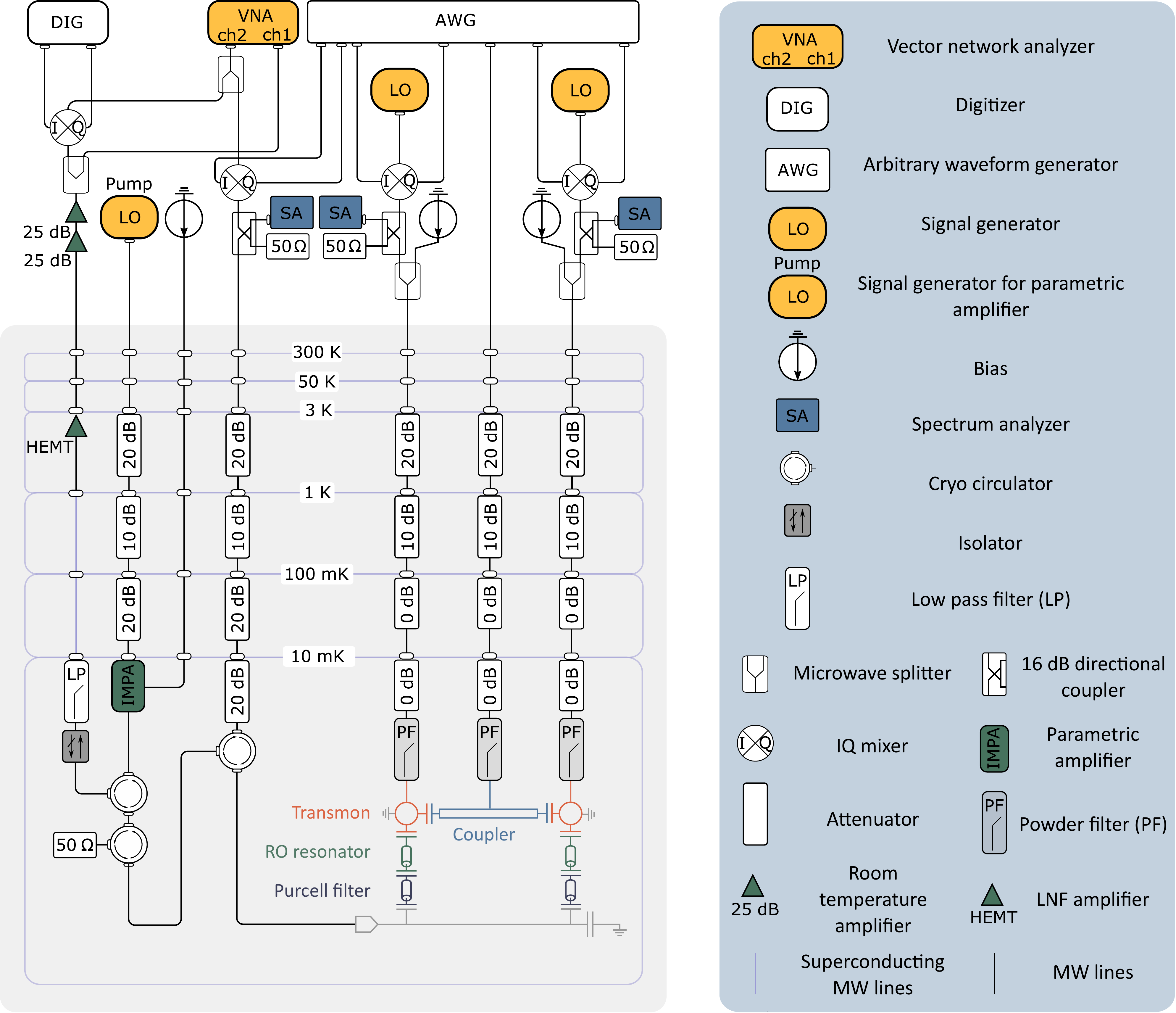}
    \caption{The scheme of the experimental setup.}
    \label{fig:experimental setup}
\end{figure*}
In preliminary measurements, we observe that coherence time $T_2^*$ decrease by half when the coupler is tuned to the gate activation flux point. At the same time relaxation time is stable with the flux change in range operating diapason. Assuming that during the gate $T_2^*$ is two times less than in Table~\ref{tab:gate params}, formula~(\ref{eq:fidelity2Q_T1T2}) yields a depolarizing error of 0.016.

Since we utilize the XEB method, it also allows us to estimate the acquired two-qubit phase, $\theta$. 
By optimizing the linear cross-entropy between the experimental data and simulated quantum circuits with an interleaved controlled-phase gate, we find $\theta = (1.03 \pm 0.03) \pi$, that cause a depolarizing error less than 0.002. 

The remaining error sources, as suggested by numerical simulations, are swap errors, meaning state population exchanges within the computational subspace, and leakage out of the computational subspace.
To estimate the contribution of leakage, we employ speckle purity benchmarking \cite{Arute2019} and extract the average state purity after executing XEB sequences.
The calculated square root of the purity is $0.955 \pm 0.001$, which corresponds to decoherence. 
Given the total depolarization error $p_2 = 0.947$, we attribute a leakage error of 0.008. 
The residual error, defined as the difference between the square root of the purity and the combined effect of $T_1$, $T_2^*$, and ZZ induced errors, is ascribed to swap errors, and is significantly small compared to decoherence error and leakage.  

To outline, we observe that coherence errors constitute the largest contribution to the two-qubit gate error, while leakage also plays a significant role. 
In contrast, for single-qubit gates, unsuppressed ZZ interactions notably degrade fidelity. 
Addressing the decrease of such errors impact, particularly the suppression of ZZ interaction, should be a priority for future research.


\section{\label{sec:Appendix3}Experimental setup}
The experimental setup is presented in Fig.~\ref{fig:experimental setup}. The device is placed in the dilution refrigerator (grey area). Transmons are coupled to a readout reflection line via a resonator and a Purcell filter. Transmon and coupler control lines are attenuated and filtered for thermalization and noise reduction, as well as the readout line. We use an impedance-matching parametric amplifier \cite{dorogov2022application} for the gain of a readout signal. Outside the refrigerator the system of microwave and dc control is assembled. For pulse generation we use IQ mixers with inputs for the sine signal generator and the arbitrary waveform generator. For the readout we use IQ mixer in the opposite direction: the amplified signal from the readout line comes to the RF port and converted to the I and Q signals, that go to the digitizer.

\newpage
\normalem{}
\bibliography{main.bib}

\end{document}